\documentclass[draftcls,10pt,onecolumn]{IEEEtran}
\IEEEoverridecommandlockouts
\usepackage{graphicx}
\usepackage{amsmath}
\usepackage{amssymb}               
\usepackage{amsfonts}               
\usepackage{amsopn,amsthm}
\usepackage{tikz}
\usetikzlibrary{arrows,shapes.symbols}
\usetikzlibrary{through}
\usetikzlibrary{fit}					
\usetikzlibrary{backgrounds,calc}	
\usepackage{color}
\allowdisplaybreaks
\begin{document}
\newtheorem{theorem}{\bf Theorem}
\newtheorem{lemma}{\bf Lemma}
\newtheorem{claim}{Claim}
\newtheorem{proposition}{\bf Proposition}
\newtheorem{conjecture}{Conjecture}
\newtheorem{corollary}{\bf Corollary}
\newtheorem{definition}{\bf Definition}
\newtheorem{example}{Example}
\newtheorem{discussion}{Discussion}
\newtheorem{remark}{\bf Remark}
\def\p{\mathsf{P}}
\def\hY{\hat{Y}}
\def\hy{\hat{y}}
\def\hz{\hat{z}}
\def\hu{\hat{u}}
\def\hZ{\hat{Z}}
\def\hS{\hat{S}}
\def\hs{\hat{s}}
\def\hR{\hat{R}}
\def\hp{\hat{P}}
\def\tp{\tilde{P}}
\def\tX{\tilde{X}}
\def\tY{\tilde{Y}}
\def\tZ{\tilde{Z}}
\def\tR{\tilde{R}}
\def\styp{\ms_{\gamma}(p\|t)}
\def\typ{\mathcal{T}_{\epsilon}^n}
\def\Reals{\mathbb R}
\def\n{\nonumber\\}
\def\be{\begin{equation}}
\def\ee{\end{equation}}
\def\bes{\begin{equation*}}
\def\ees{\end{equation*}}
\def\beq{\begin{eqnarray}}
\def\eeq{\end{eqnarray}}
\def\beqs{\begin{eqnarray*}}
\def\eeqs{\end{eqnarray*}}
\def\ma{{\mathcal A}}
\def\mb{{\mathcal B}}
\def\mc{{\mathcal C}}
\def\md{{\mathcal D}}
\def\me{{\mathcal E}}
\def\mf{{\mathcal F}}
\def\mg{{\mathcal G}}
\def\mh{{\mathcal H}}
\def\mi{{\mathcal I}}
\def\mj{{\mathcal J}}
\def\mk{{\mathcal K}}
\def\ml{{\mathcal L}}
\def\mm{{\mathcal M}}
\def\mn{{\mathcal N}}
\def\mo{{\mathcal O}}
\def\mp{{\mathcal P}}
\def\mq{{\mathcal Q}}
\def\mr{{\mathcal R}}
\def\ms{{\mathcal S}}
\def\mt{{\mathcal T}}
\def\mv{{\mathcal V}}
\def\mw{{\mathcal W}}
\def\mx{{\mathcal X}}
\def\my{{\mathcal Y}}
\def\mz{{\mathcal Z}}
\def\ua{\mathbf{a}}
\def\ub{\mathbf{b}}
\def\uc{\mathbf{c}}
\def\ud{\mathbf{d}}
\def\ue{\mathbf{e}}
\def\uf{\mathbf{f}}
\def\ug{\mathbf{g}}
\def\uh{\mathbf{h}}
\def\ui{\mathbf{i}}
\def\uj{\mathbf{j}}
\def\uk{\mathbf{k}}
\def\ul{\mathbf{l}}
\def\um{\mathbf{m}}
\def\un{\mathbf{n}}
\def\uo{\mathbf{o}}
\def\br{\mathbf{R}}
\def\bC{\mathbf{C}}
\def\F{\mathbf{F}}
\def\bp{\mathbf{P}}
\def\uq{\mathbf{q}}
\def\ur{\mathbf{r}}
\def\us{\mathbf{s}}
\def\ut{\mathbf{t}}
\def\uu{\mathbf{u}}
\def\uv{\mathbf{v}}
\def\uw{\mathbf{w}}
\def\e{\mathbb{E}}
\def\tt{\mathbf{1}((u^n,v^n,w^n,y^n,z^n)\in\styp)}
\def\tv#1{\left\|#1\right\|_1}
\def\apx#1{\stackrel{#1}{\approx}}
\def\ux{x^n}
\def\uy{y^n}
\def\uz{z^n}
\def\uyy{\mathbf{y}}

\def\uhy{\mathbf{\hat{y}}}
\def\vanish{\xrightarrow{\footnotesize exp} 0}
\def\mathllap{\mathpalette\mathllapinternal}
 \def\mathllapinternal#1#2{%
 \llap{$\mathsurround=0pt#1{#2}$}
 }
 \def\clap#1{\hbox to 0pt{\hss#1\hss}}
 \def\mathclap{\mathpalette\mathclapinternal}
 \def\mathclapinternal#1#2{%
 \clap{$\mathsurround=0pt#1{#2}$}%
 }
 \def\mathrlap{\mathpalette\mathrlapinternal}
 \def\mathrlapinternal#1#2{%
 \rlap{$\mathsurround=0pt#1{#2}$}
 }
\def\Sum{\ensuremath\mathop{\scalebox{1.2}{$\displaystyle\sum$}}}
\def\Prod{\ensuremath\mathop{\scalebox{1.2}{$\displaystyle\prod$}}}
\def\im{\imath}
\def\s#1{\mathsf{#1}}
\def\ep{\epsilon}
\def\Var{\mathsf{Var}}
\def\TV{\mathsf{TV}}
\def\sec{\mathsf{sec}}
\def\Cov{\mathsf{Cov}}
\def\SW{\mathsf{SW}}
\def\ap{\mathsf{Apx}}
\def\sm{\mathsf{M}}
\def\sf{\mathsf{F}}
\def\sj{\mathsf{J}}
\def\enc{\mathsf{Enc}}
\def\dec{\mathsf{Dec}}
\sloppy

\title{A Technique for Deriving One-Shot  Achievability Results in Network Information Theory}

\author{Mohammad~Hossein~Yassaee, Mohammad~Reza~Aref, Amin~Gohari\\Information Systems and Security Lab (ISSL),\\Sharif University of Technology, Tehran, Iran,\\
E-mail: yassaee@ee.sharif.edu, \{aref,aminzadeh\}@sharif.edu.
 \thanks{ This work was supported by Iran-NSF under grant No. 88114.46.}}





\maketitle

\begin{abstract}
This paper proposes a novel technique to prove a one-shot version of achievability results in network information theory. The technique is not based on covering and packing lemmas. In this technique, we use an stochastic encoder and decoder with a particular structure for coding that resembles both the ML and the joint-typicality coders. Although stochastic encoders and decoders do not usually enhance the capacity region, their use simplifies the analysis. The Jensen inequality lies at the heart of error analysis, which enables us to deal with the expectation of many terms coming from stochastic encoders and decoders at once. The technique is illustrated via several examples: point-to-point channel coding, Gelfand-Pinsker, Broadcast channel (Marton), Berger-Tung, Heegard-Berger/Kaspi, Multiple description coding and Joint source-channel coding over a MAC. Most of our one-shot results are new. The asymptotic forms of these expressions is the same as that of classical results. Our one-shot bounds in conjunction with multi-dimensional Berry-Essen CLT  imply new results in the finite blocklength regime. In particular applying the one-shot result for the memoryless broadcast channel in the asymptotic case, we get the entire region of Marton's inner bound without any need for time-sharing. 
\end{abstract}

\section{Introduction}
Information theory aims to find optimal reliable communication rates in networks. The combinatorial structure of networks makes the problem difficult in general. However one can employ law of large numbers by looking at asymptotic behavior of networks for large blocklengths. But this comes at the cost of a long delay. This motivates looking at the problem in the so called ``finite blocklength regime." The blocklength in this regime is not infinitely long, but is sufficiently large for certain CLTs to hold. Originally studied by Strassen \cite{strassen}, there has been a recent surge of works on this topic following the results of Polyanskiy et al \cite{PPV} (see for instance \cite{verdujscc}-\cite{Tan}). 

In this paper we consider \emph{one-shot} network information theory where a \emph{single} use of the network is allowed. In this case the probability of error cannot  necessarily be driven to zero. Further, well-known techniques such as joint typicality and time sharing are not applicable here. Given an admissible probability of error, our goal is to find a characterization of a set of achievable rates that resembles the form of the asymptotic results. There has been some previous work along this direction. Wang and Renner \cite{Renner1} derive one-shot upper and lower bounds for the problem of transmission of classical information over a classic-quantum channel (see also \cite{Renner2}). 
 Recently Verdu has proposed a one-shot version of the covering and packing lemmas, and has applied it to a set of classical problems in information theory \cite{verdual}. 

Our main contribution is a proof technique for deriving the results on the one-shot region. The technique uses elementary tools and is not based on extensions of packing or covering lemmas. It is based on a particular construction for  encoder and decoders that is not ML, but resembles both the ML and jointly typical coders. Our proposed decoders are stochastic and intuitively attempt to pass the received symbol through a certain inverse conditional distribution. The Jensen's inequality is central to the analysis of the error. The technique can be widely applied to problems of network information theory. To illustrate this, we derive new results for the problems of Gelfand-Pinsker, broadcast channel, joint-source channel coding over MAC, Berger-Tung, Heegard-Berger/Kaspi and multiple description coding. The asymptotic forms of these expressions is the same as that of classical results. Our one-shot bounds also imply new results in the finite blocklength regime. 

The most related previous work is that of Verdu \cite{verdual}. Whereas \cite{verdual} proposes a one-shot covering and packing lemmas to solve network problems, we propose a direct analysis comprising of a chain of inequalities. By bypassing the need for covering and packing lemmas, we can provide bounds for problems that were originally solved using mutual covering and packing lemmas in the asymptotic regime. This is helpful because no one-shot extension of the mutual covering and packing lemma exists. We discuss this point in more details in Remark \ref{remarkM}.

This paper is organized as follows: In Section \ref{prelim} we provide some definitions. This is followed by three sections that provide application of the technique to different scenarios. In Section 
\ref{MTCC} we consider three problems of multi-terminal channel coding, namely, point-to-point channel, Gelfand-Pinsker and broadcast channel (Marton). 
In Section 
\ref{MTLSC} we consider three problems of lossy multi-terminal source coding, i.e. Berger-Tung, Heegard-Berger/Kaspi and Multiple description coding. Lastly in Section \ref{JSCC} we study a joint source-channel coding problem of transmission correlated sources over a MAC. In each of these problems we provide a one-shot achievability result. Corresponding finite blocklength results could be derived from these results. To illustrate this, we have derived such bounds for the Gelfand-Pinsker and broadcast channel problems.

\section{Definitions}
\begin{definition}
Given a pmf $p_{X,Y}$, the conditional information of $x$ given $y$ is defined by
\[h_p(x|y)=\log\frac{1}{p_{X|Y}(x|y)}.\]
\end{definition}
\begin{definition}
For a pmf $p_{X,Y,Z}$, the \emph{conditional information density} $\im(x;y|z)$ is defined by
\[ \im_p(x;y|z):=\log \frac{p(x,y|z)}{p(x|z)p(y|z)},\]
and for general r.v.'s it is defined by
\[ \im_p(x;y|z):=\log \frac{\mathsf{d} p_{X,Y|Z}}{\mathsf{d}(p_{X|Z}\times p_{Y|Z})}(x,y,z).\]
Whenever the underlying distribution is clear from the context, we drop the subscript $p$ from $\im_p(x;y|z)$.
\end{definition}
\begin{definition}
Let $\mathbf{X}$ be a multi-dimensional normal variable with zero mean and covariance matrix $\mathsf{V}$. The complementary multivariate Gaussian cumulative distribution region associated with $\mathsf{V}$ is defined by
 \[
\mq^{-1}(\s{V},\epsilon):=\{\mathbf{x}:\s{P}(\mathbf{X\le x})\ge 1-\ep\}.
\]
\end{definition}

We use $\sm$ and $\sj$ to denote size of alphabets of random variables $M$ and $J$, respectively, i.e. $\sm=|\mathcal{M}|$ and $\sj=|\mathcal{J}|$. All the logarithms are in base two throughout this paper.
\label{prelim}

\section{Multi-terminal channel coding problems}
\label{MTCC}
To illustrate the application of our technique to multi-terminal channel coding problems, we study the problems of point-to-point channel, Gelfand-Pinsker and Broadcast channels (Marton) in this section.

\subsection{Point-to-point channel}
We begin our illustration of the one-shot achievability proof with the classical point-to-point channel. Consider a channel with the law $q_{Y|X}$ and an input distribution $q_X$. Let $\mc=\{X(1),\cdots,X(\sm)\}$ be a random codebook where the elements $X(i)$ are drawn independently from $q_X$ (each codeword $X(i)$ is only a single rv). As usual, $X(m)$ is the codeword used for transmission of the message $m$. For the decoding we use an stochastic variation of MAP decoding. Instead of declaring the message $\hat{m}$ with maximal posterior probability as in MAP, the decoder randomly draws a message $\hat{m}$ from the conditional pmf $P_{M|Y}$, where $P$ is the induced pmf by the code, $P_{M,Y}(m,y)=\frac{1}{\sm}q(y|X(m))$.\footnote{The pmf is random due to the random codebook.} More specifically,
\be
P_{M|Y}(\hat{m}|y)=\dfrac{q(y|X(\hat{m}))}{\sum_{\bar{m}}q(y|X(\bar{m}))}=\dfrac{2^{\im_q(y;X(\hat{m}))}}{\sum_{\bar{m}}2^{\im_q(y;X(\bar{m}))}}.
\ee 
The mutual information term $\im_q(y;X(\hat{m}))$ is computed using pmf $q_Xq_{Y|X}$ that has nothing to do with the pmf induced by the code. However the sequence $X(\hat{m})$ itself is random, hence we have used $P_{M|Y}(\hat{m}|y)$ (capital $P$) to denote the pmf.

 We refer this decoder as \emph{stochastic likelihood coder} (SLC), or as \emph{stochastic mutual information coder} (SMC).\footnote{The reason for introducing two names for apparently the same object will become clear later. These decoders will not be the same in other problems.} The second equality shows that the probability of selecting a message is proportional to two to the power of its mutual information with the received output. So codewords with higher mutual information have a higher chance of being selected as the output of the decoder. This resembles the widely used joint typicality decoder in the asymptotic regime. 
\begin{theorem}
The \emph{expected value} of the probability of \emph{correct decoding} of SLC (or SMC) for a randomly generated codebook of size $\sm$ is bounded from below by
\end{theorem}     
\begin{equation}
\e_{\mc} \mathsf{P}[C]\ge\e_{q_{XY}} \dfrac{1}{1+(\sm-1)2^{-\im_q(X;Y)}}.
\end{equation}
\begin{IEEEproof}
Observe that the joint distribution of random variables factors as,
\bes
P_{MY\hat{M}}(m,y,\hat{m})=\frac{1}{\sm}q(y|X(m))P_{M|Y}(\hat{m}|y),
\ees
and the probability of correct decoding can be written as $\mathsf{P}[C]=\sum_{m,y}P_{MY\hat{M}}(m,y,m)$, hence we have:
\begin{IEEEeqnarray}{lCl}
\e\mathsf{P}[C]&=&\e\sum_{m,y}\frac{1}{\sm}q(y|X(m))\dfrac{2^{\im_q(y;X({m}))}}{\sum_{\bar{m}}2^{\im_q(y;X(\bar{m}))}}\\
                         &=&\e\sum_{y}q(y|X(1))\dfrac{2^{\im_q(y;X({1}))}}{\sum_{\bar{m}}2^{\im_q(y;X(\bar{m}))}}\\
                         &=&\sum_{y}\e_{X(1)}\e_{\mathcal{C}|X(1)}q(y|X(1))\dfrac{2^{\im_q(y;X({1}))}}{\sum_{\bar{m}}2^{\im_q(y;X(\bar{m}))}}\label{eq:p2p1}\\
                         &\ge&\sum_{y}\e_{X(1)}q(y|X(1))\dfrac{2^{\im_q(y;X({1}))}}{\e_{\mathcal{C}|X(1)}\sum_{\bar{m}}2^{\im_q(y;X(\bar{m}))}}\label{eq:p2p2}\\
                         &=&\sum_{y}\e_{X(1)}q(y|X(1))\dfrac{2^{\im_q(y;X({1}))}}{2^{\im_q(y;X({1}))}+(\sm-1)}\label{eq:p2p3}\\
                         &=&\sum_{x,y}q(x)q(y|x)\dfrac{2^{\im_q(y;x)}}{2^{\im_q(y;x)}+(\sm-1)}\label{eq:p2p4}\\
                         &=&\e_{XY} \dfrac{1}{1+(\sm-1)2^{-\im_q(X;Y)}},
\end{IEEEeqnarray}
where \eqref{eq:p2p1} follows from the rule of iterated expectation, \eqref{eq:p2p2} follows from the Jensen inequality for the convex function $f(x)=\dfrac{1}{x}$ on $\mathbb{R}_+$, and \eqref{eq:p2p3} follows from the following equation for $\bar{m}\neq1$,
\begin{IEEEeqnarray*}{lCl}
\e_{\mathcal{C}|X(1)}2^{\im(y;X(\bar{m}))}=\sum_{x}q(x)2^{\im(y;x)}=\sum_{x}q(x|y)=1,
\end{IEEEeqnarray*}
where we use the fact that $X(\bar{m})$ is independent of $X(1)$ for $\bar{m}\neq1$ and is drawn from $q_X$.
\end{IEEEproof}
\subsection{Gelfand-Pinsker}
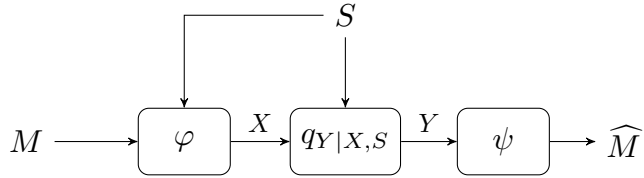
\begin{figure}\begin{center} \begin{tikzpicture}[scale=1.4,>=
   stealth'
  ]
\def\bscale{1.5}
\tikzstyle{enc}=[scale=.8,draw=black, minimum width=2.9em, rounded corners,
    text centered, minimum height=2.1em
    ]
    \tikzstyle{ch}=[scale=.8,draw=black, minimum width=2.6em, rounded corners,
    text centered, minimum height=6em,draw=blue,
    ]
\tikzstyle{ann} = [above, text width=2em]
\tikzstyle{dec} = [enc, text width=2.4em, 
    minimum height=2em,fill=gray!70!white,draw=black]
       \node (e0) [enc,scale=\bscale]{$q_{Y|X,S}$}; 
          \path (e0)+(1.5,0) node (d1) [enc,scale=\bscale]{$\psi$}; 
                    \path (e0)+(0,1.2) node (st)[scale=1.2] {$S$}; 
     \path (e0.180)+(-1,0) node (dec) [enc,scale=\bscale]{$\varphi$}; 
    \path (d1.east)+(.7,0) node (d10) [scale=1.2]{$\widehat{M}$};
          \path (dec)+(-1.5,0) node (es) [scale=1.2]{${M}$};
  \path[draw,->] (e0)-- node(m)[above]{$Y$}(d1);
      \path[draw,<-] (e0)-- node[above]{$X$}(dec);
          \draw[->](d1)--(d10);
           \draw[<-](dec)--(es);
           \draw[->] (st)--(e0);
           \draw[->] (st)-|(dec);
  \end{tikzpicture}
  \caption{Coding over a state-dependent channel.}\label{fig:gp}
 \end{center}

  \end{figure}
  
Consider the problem of transmitting a message over a state-dependent channel with state information available at the encoder, depicted in Fig. \ref{fig:gp}. Let $q_S$ and $q_{Y|X,S}$ be the state's pmf  and  the channel transition probability, respectively.
\subsubsection{One-shot achievability}
\begin{definition}
 An $\sm$-code for state-dependent channel consists of a (possibly stochastic) encoder $\varphi:[1:\sm]\times\ms\mapsto\mx$, and  a (possibly stochastic) decoder $\psi:\my\mapsto[1:\sm]$. 
 \end{definition}
 \begin{theorem}
Given any $q_{U|S}$ and function $x(u,s)$, there is an $\sm$-code for a single use of the channel whose probability of correct decoding is bounded from below by
\begin{equation}\label{eq:GP}
\e_{USY} \dfrac{1}{(1+\sj^{-1}2^{\im_q(U;S)})(1+\sm\sj2^{-\im_q(U;Y)})},
\end{equation}
where $\sj>0$ is an arbitrary integer and $q(u,s,x,y)=q(s)q(u|s)\mathbf{1}[x=x(u,s)]q(y|x,s)$. Moreover, loosening this bound gives the following upper bound on the error probability of the code,
\begin{align}\p[\log\sj&-\im_q(U;S)<\gamma\ ,\ \mathsf{or}\n& \im_q(U;Y)-\log\sm\sj<\gamma]+3\times2^{-\gamma},\label{eq:100}\end{align}
where $\gamma$ is any positive number.
\end{theorem} 
\begin{remark} If we apply the above result to $n$ copies of a memoryless state dependent channel, we recover the asymptotic Gelfand-Pinsker result. In this derivation 
the first term in the denominator of \eqref{eq:GP}, $	1+\sj^{-1}2^{\im(U;S)}$ corresponds to a covering lemma in the asymptotic case, while the second term $1+\sm\sj2^{-\im(U;Y)}$ corresponds to a packing lemma. Observe that the first term is proportional to $\sj^{-1}$ whereas the second term is proportional to $\sm\sj$. Thus the above formula combines covering and packing lemmas at once.
\end{remark}
\begin{remark}
If we further loosen the first term of eq. \eqref{eq:100} using the union bound, we get Verdu's bound on this problem \cite{verdual} except for the term $3\times2^{-\gamma}$. This residual term is not of significance in the finite blocklength $n$-letter regime where we choose $\gamma$ of the order of $\log(n)$ (see Theorem \ref{thm:GPFB}); the main contribution comes from the probability terms. In a concurrent work \cite{watanabe}, Watanabe et.al., prove an expression similar to eq. \eqref{eq:100} using a different method based on channel simulation. They also applied their approach to the problem of source coding with a helper and to the Wyner-Ziv problem. It is not clear whether their approach is applicable to the scenarios such as broadcast channel, multiple description coding, etc that is solved in the asymptotic case using the multivariate covering lemma, since no extension of channel simulation (based on the work of Cuff \cite{cuff12}) is known for multiuser scenarios. Nonetheless, our technique bypasses the need for either an extension of covering lemma to multivariate covering, or a multi-terminal extension of the channel simulation result. See also Remark \ref{remarkM}.  
\end{remark}
\begin{IEEEproof}

 Let $\mc=\{U(m,j)\}_{m=1,j=1}^{\sm,\sj}$ be a random codebook whose elements are drawn independently from $q_U$. Here $J$ is introducing redundancy but since it will be decoded at the receiver we can view it as a dummy message.

\subsubsection*{Encoding} Instead of using conventional random covering, we use an SMC which acts as follows. Given $m$ and $s$, the SMC chooses an index $j$ with the probability
\[
P_{\enc}(j|m,s)=\dfrac{2^{\im_q(s;U(m,j))}}{\sum_{\tilde{j}}2^{\im_q(s;U(m,\tilde{j}))}}.
\]  
Then the encoder transmits $x(U(m,j),s)$ through the channel. Observe that the above SMC resembles a joint-typical encoder of the asymptotic regime. Given $m$ and $s$, the higher the information between $U(m,j)$ and $s$, the more likely we choose it at the encoder.
\subsubsection*{Decoding} In contrast to the point-to-point problem, computing the error probability of SLC is challenging. An SLC uses 
the induced $P_{M,J|Y}$ by the code. Instead, we use an SMC with the following rule for decoding. Observing $y$, decoder uses the following SMC to find both the message $m$ and the dummy message $j$:
 \[
P_{\dec}(\hat{m},\hat{j}|y)=\dfrac{2^{\im_q(y;U(\hat{m},\hat{j}))}}{\sum_{\bar{m},\bar{j}}2^{\im_q(s;U(\bar{m},\bar{j}))}}.
\] 
\subsubsection*{Analysis} We declare an error if $(\hat{m},\hat{j})\neq (m,j)$. Observe that the joint distribution of random variables factors as,
\begin{align*}
P_{MJSY\hat{M}\hat{J}}(m,j,s,y,\hat{m},\hat{j})=\frac{1}{\sm}q(s)P_{\enc}(j|m,s)q(y|U(m,j),s)P_{\dec}(\hat{m},\hat{j}|y),
\end{align*}
 where $q(y|U(m,j),s)=q_{Y|X,S}\big(y|x(U(m,j),s),s\big)$. The probability of correct decoding is $\mathsf{P}[C]=\sum_{m,j,s,y}P_{MJSY\hat{M}\hat{J}}(m,j,s,y,m,j)$; hence we have:
\begin{IEEEeqnarray}{rCl}
\e\mathsf{P}&[C]&=\e\sum_{m,j,s,y}\frac{1}{\sm}q(s)\dfrac{2^{\im(s;U(m,j))}}{\sum_{\tilde{j}}2^{\im(s;U(m,\tilde{j}))}}q(y|U(m,j),s)\dfrac{2^{\im(y;U(m,j))}}{\sum_{\bar{m},\bar{j}}2^{\im(y;U(\bar{m},\bar{j}))}}\\
                         &=&\e\sum_{s,y}\sj q(s)\dfrac{2^{\im(s;U(1,1))}}{\sum_{\tilde{j}}2^{\im(s;U(1,\tilde{j}))}}q(y|U(1,1),s)\dfrac{2^{\im(y;U(1,1))}}{\sum_{\bar{m},\bar{j}}2^{\im(y;U(\bar{m},\bar{j}))}}\label{eq:101}\\
                         &\ge&\sum_{s,y}\textcolor{blue!90!black}{\e_{U(1,1)}}\left(\dfrac{\textcolor{blue!90!black}{\sj q(s)2^{\im(s;U(1,1))}}}{\textcolor{red!60!blue}{\e_{\mc|U(1,1)}\sum_{\tilde{j}}2^{\im(s;U(1,\tilde{j}))}}} \textcolor{blue!90!black}{q(y|U(1,1),s)}\dfrac{\textcolor{blue!90!black}{2^{\im(y;U(1,1))}}}{\textcolor{red!60!blue}{\e_{\mc|U(1,1)}\sum_{\bar{m},\bar{j}}2^{\im(y;U(\bar{m},\bar{j}))}}}\right)\IEEEeqnarraynumspace\label{eq:102}\\
                         &\ge&\sum_{s,y}\e_{U(1,1)}\left(\dfrac{\sj q(s)2^{\im(s;U(1,1))}}{2^{\im(s;U(1,1))}+\sj} q(y|U(1,1),s)\dfrac{2^{\im(y;U(1,1))}}{2^{\im(y;U(1,1))}+\sm\sj}\right)\label{eq:103}\\
                         &=&\sum_{u,s,y}q(u,s,y)\dfrac{\sj}{2^{\im(s;u)}+\sj}\cdot\dfrac{2^{\im(y;u)}}{2^{\im(y;u)}+\sm\sj}\label{eq:104}\\
                         &=&\e_{USY} \dfrac{1}{(1+\sj^{-1}2^{\im(U;S)})(1+\sm\sj2^{-\im(U;Y)})},\label{eq:105}
\end{IEEEeqnarray}
where \eqref{eq:101} is due to symmetry, the \underline{main step} \eqref{eq:102} follows from Jensen inequality for the two-valued convex function $f(x_1,x_2)=\dfrac{1}{x_1x_2}$ on $\mathbb{R}^2_+$, \eqref{eq:103} follows from the fact that $U(i,j)$ is independent of $U(1,1)$ for $(i,j)\neq(1,1)$ and generated according to $q_U$, and \eqref{eq:104} follows from the fact that $U(1,1)$ is  generated according to $q_U$.
\subsubsection*{Deriving the loosened bound \eqref{eq:100}}
\begin{align}
\e_{USY} \dfrac{1}{(1+\sj^{-1}2^{\im(U;S)})(1+\sm\sj2^{-\im(U;Y)})}&\ge\e_{USY} \dfrac{\mathbf{1}\left\{\log\sj-\im(U;S)\ge\gamma,\ \im(U;Y)-\log\sm\sj\ge\gamma\right\}}{(1+\sj^{-1}2^{\im(U;S)})(1+\sm\sj2^{-\im(U;Y)})}\\
                                                                                                         &\ge\dfrac{1}{(1+2^{-\gamma})^2}\mathsf{P}\left[\log\sj-\im(U;S)\ge\gamma,\ \im(U;Y)-\log\sm\sj\ge\gamma\right]
\end{align}
\begin{align}
\mathsf{P}[\me]&\le1-\dfrac{1}{(1+2^{-\gamma})^2}\mathsf{P}\left[\log\sj-\im(U;S)\ge\gamma,\ \im(U;Y)-\log\sm\sj\ge\gamma\right]\\
                         &=\mathsf{P}\left[\log\sj-\im(U;S)<\gamma,\ \mathsf{or}\ \im(U;Y)-\log\sm\sj<\gamma\right]\n
                         &\qquad\qquad\qquad\qquad+\left(1-\dfrac{1}{(1+2^{-\gamma})^2}\right)\mathsf{P}\left[\log\sj-\im(U;S)\ge\gamma,\ \im(U;Y)-\log\sm\sj\ge\gamma\right]\\
                         &\le\mathsf{P}\left[\log\sj-\im(U;S)<\gamma,\ \mathsf{or}\ \im(U;Y)-\log\sm\sj<\gamma\right]+\left(1-\dfrac{1}{(1+2^{-\gamma})^2}\right)\\
                         &\le\mathsf{P}\left[\log\sj-\im(U;S)<\gamma,\ \mathsf{or}\ \im(U;Y)-\log\sm\sj<\gamma\right]+3\times2^{-\gamma}.
\end{align}
\end{IEEEproof}
\subsubsection{Second Order achievability of Gelfand-Pinsker channel}
\begin{theorem}\label{thm:GPFB}
Given a memoryless state-dependent channel $(q_S,q_{Y|X,S})$ with state known non-causally at the encoder, for any $(q_{U|S},x(u,s))$, the following rate is $(n,\epsilon)$-achievable:
\be
R=I(U;Y)-I(U;S)-\frac{1}{\sqrt{n}}R_D-O\left(\dfrac{\log n}{n}\right)
\ee
where
\begin{equation}
R_D=\min_{R:\exists \tilde{R}, \ \mathrm{s.t.}\  [\tilde{R},R-\tilde{R}]^{\mathsf{T}}\in \mq^{-1}(\mathbb{V}_{\mathsf{GP}},\ep)} R,
\end{equation}
and

\be
\mathbb{V}_{\mathsf{GP}}=\mathsf{Cov}\left[\begin{matrix}\im(U;S)\\ \im(U;Y)\end{matrix}\right].
\end{equation}
\end{theorem}
\begin{IEEEproof}
We apply \eqref{eq:100} to $n$ use of the channel. Assume that $q_{U^n}(u^n)=\prod_{i=1}^nq_U(u_i)$, so $(U^n,S^n,Y^n)$ are i.i.d.. Substituting $\gamma=\dfrac{1}{2}\log n$ in \eqref{eq:100} implies: 
\begin{align}
P[\me]\leq \p\left(\log\sj-\im(U^n;S^n)\leq \dfrac{1}{2}\log n,~~ \mathsf{or}~~\im(U^n;Y^n)-\log \sm\sj\leq \dfrac{1}{2}\log n\right)+\dfrac{3}{\sqrt{n}}.
\end{align}
Given $\epsilon>0$, finding a code such that for some $\epsilon'\leq \epsilon$ 
\begin{align}
\epsilon'=\p\left(\log\sj-\im(U^n;S^n)\leq \dfrac{1}{2}\log n,~~ \mathsf{or}~~\im(U^n;Y^n)-\log \sm\sj\leq \dfrac{1}{2}\log n\right)+\dfrac{3}{\sqrt{n}}.
\end{align}
implies that $\epsilon$ is an upper bound on $P[\me]$. Equivalently, we would like to find a code such that
\begin{align}
1-\ep'-\dfrac{3}{\sqrt{n}}&=\p\left(\log\sj-\im(U^n;S^n)>\dfrac{1}{2}\log n,~~~~~\im(U^n;Y^n)-\log \sm\sj>\dfrac{1}{2}\log n\right).\label{eq:110}
\end{align}
Let $\log\sj=nI(U;S)+\sqrt{n}\tR+\dfrac{1}{2}\log n$ and $\log\sm=n(I(U;Y)-I(U;S))-\sqrt{n}R-\log n$. The random variables $\im(U^n;S^n)$ and $\im(U^n;Y^n)$ are  sum of i.i.d. random variabels. Applying multi-dimensional Berry-Essen CLT \cite{CLT} to \eqref{eq:110} implies the following equivalent form:
\begin{align}
1-\ep'-O(\dfrac{1}{\sqrt{n}})=\p_G\left(\left[\begin{array}{l}G_1\\G_2\end{array}\right]\le\left[\begin{array}{c}\tR\\R-\tR\end{array}\right]\right),
\end{align}
where $G=[G_1\ G_2]^{\mathsf{T}}$ is a multidimensional normal r.v. with zero mean and $\Cov G=\mathbb{V}_{\mathsf{GP}}$. Using the definition of $\mq^{-1}(\mathsf{V},\ep)$ and smoothness of distribution of normal r.v., we get 
\be [\tilde{R},R-\tilde{R}]^{\mathsf{T}}\in \mq^{-1}(\mathbb{V}_{\mathsf{GP}},\ep')+O\left(\dfrac{\log n}{\sqrt{n}}\right).\ee
Taking the limit of $\ep'\uparrow\ep$, we see that for any arbitrary $ [\tilde{R},R-\tilde{R}]^{\mathsf{T}}$ in $\mq^{-1}(\mathbb{V}_{\mathsf{GP}},\ep)$, we can achieve $P[\me]\leq \ep$. This completes the proof.
\end{IEEEproof}

\subsection{Broadcast channel}
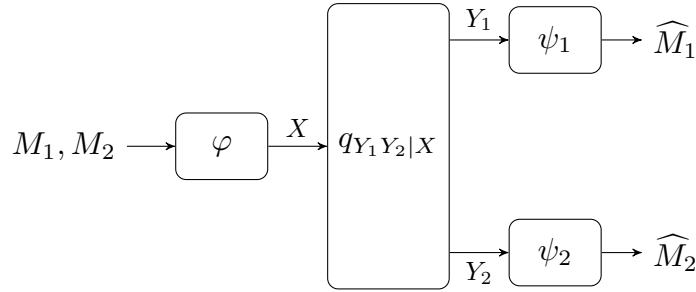
\begin{figure}  \begin{center}\begin{tikzpicture}[scale=1.4,>=
   stealth'
  ]
\def\bscale{1.5}
\tikzstyle{enc}=[scale=.8,draw=black, minimum width=2.9em, rounded corners,
    text centered, minimum height=2.1em
    ]
    \tikzstyle{ch}=[scale=.8,draw=black, minimum width=2.6em, rounded corners,
    text centered, minimum height=6em,draw=blue,
    ]
\tikzstyle{ann} = [above, text width=2em]
\tikzstyle{dec} = [enc, text width=2.4em, 
    minimum height=2em,fill=gray!70!white,draw=black]
       \node (e0) [enc,scale=\bscale,minimum height=9em]{$q_{Y_1Y_2|X}$}; 
          \path (e0.60)+(1,0) node (d1) [enc,scale=\bscale]{$\psi_1$}; 
          \path (e0.-60)+(1,0) node (d2) [enc,scale=\bscale]{$\psi_2$}; 
     \path (e0.180)+(-1,0) node (dec) [enc,scale=\bscale]{$\varphi$}; 
    \path (d1.east)+(.7,0) node (d10) [scale=1.2]{$\widehat{M}_1$};
     \path (d2.east)+(.7,0) node (d20) [scale=1.2]{$\widehat{M}_2$};
          \path (dec)+(-1.5,0) node (es) [scale=1.2]{${M}_1,{M}_2$};
  \path[draw,->] (e0.-300)-- node(m)[above]{$Y_1$}(d1);
    \path[draw,->] (e0.300)-- node(m2)[below]{$Y_2$}(d2);
      \path[draw,<-] (e0)-- node[above]{$X$}(dec);
          \draw[->](d1)--(d10);
           \draw[->](d2)--(d20);
           \draw[<-](dec)--(es);
  \end{tikzpicture}
  \caption{Coding over a broadcast channel.}\label{fig:bc}
 \end{center}

  \end{figure}
Consider the problem of transmission of private messages over a broadcast channel depicted in Fig. \ref{fig:bc}. Let $q_{Y_1Y_2|X}$ be the channel transition probability.
\subsubsection{One-shot achievability}
\begin{definition}
 An $(\sm_0,\sm_1,\sm_2)$-code for broadcast channel consists of a (possibly stochastic) encoder $\varphi:[1:\sm_0]\times[1:\sm_1]\times[1:\sm_2]\mapsto\mx$, and  two (possibly stochastic) decoders $\psi_k:\my_k\mapsto[1:\sm_0]\times[1:\sm_k], k=1,2$. 
 \end{definition}
 
 \begin{theorem}
Given any $q_{U_0,U_1,U_2}$ and function $x(u_0,u_1,u_2)$, there is an $(\sm_0,\sm_1,\sm_2)$-code for a single use of the channel whose probability of correct decoding is bounded from below by
\begin{align*}
\e\bigg[\left(1+(\sj_1\sj_2)^{-1}2^{\im(U_1;U_2|U_0)}\right)\prod_{k=1}^2(1+\sm_k\sj_k2^{-\im(U_k;Y_k|U_0)}+\sm_0\sj_k\sm_k2^{-\im(U_0U_k;Y_k)})\bigg]^{-1},
\end{align*}
where $\sj_1,\sj_2>0$ are arbitrary integers. Moreover, loosening this bound gives the following upper bound on error probability of the code,
\begin{align}\p\big[\log&\sj_1\sj_2-\im(U_1;U_2|U_0)<\gamma\ ,\qquad \mathsf{or}\n& \im(U_1;Y_1|U_0)-\log\sm_1\sj_1<\gamma\ ~~,\mathsf{or}~~ \im(U_0U_1;Y_1)-\log\sm_0\sm_1\sj_1<\gamma\ \mathsf{or}\n& \im(U_2;Y_2|U_0)-\log\sm_2\sj_2<\gamma~~,\mathsf{or}~~ \im(U_0U_2;Y_2)-\log\sm_0\sm_2\sj_2<\gamma\big]+17\times2^{-\gamma},\label{eq:200-b}\end{align}
where $\gamma$ is any positive number.
\end{theorem} 
\begin{corollary}
Given any $q_{U_1,U_2}$ and function $x(u_1,u_2)$, there is an $(\sm_1,\sm_2)$-code for a single use of the channel whose probability of correct decoding is bounded from below by
\begin{align*}
\e\bigg[\left(1+(\sj_1\sj_2)^{-1}2^{\im(U_1;U_2)}\right)\prod_{k=1}^2(1+\sm_k\sj_k2^{-\im(U_k;Y_k)})\bigg]^{-1},
\end{align*}
where $\sj_1,\sj_2>0$ are arbitrary integers. Moreover, loosening this bound gives the following upper bound on error probability of the code,
\begin{align}\p\big[\log&\sj_1\sj_2-\im(U_1;U_2)<\gamma\ ,\qquad \mathsf{or}\n& \im(U_1;Y_1)-\log\sm_1\sj_1<\gamma\ ,\ \mathsf{or}\n& \im(U_2;Y_2)-\log\sm_2\sj_2<\gamma\big]+7\times2^{-\gamma},\label{eq:200}\end{align}
where $\gamma$ is any positive number.
\end{corollary} 

\begin{remark}\label{remarkM}
Verdu derives a one-shot bound for the same problem in \cite{verdual}. He derives the bound by proposing a one-shot covering and packing lemma. However to get access to the boundary of Marton's inner bound one needs a mutual covering lemma (since time sharing is not possible in one-shot and not useful in finite block length regime). For this reason Verdu's result seems to be weaker than ours. Our technique allows us to bypass the need for developing a one-shot version of the mutual covering lemma. 
\end{remark}

\begin{IEEEproof}
For simplicity we prove the one-shot version of Marton with two auxiliaries (we will put the full proof in the next version of this draft). 
We only show the lower bound on probability of correct decoding. Derivation of the loosened bound is similar to that of Gelfand-Pinsker and thus omitted. 
\subsubsection*{Random codebook generation}Let $\mc=\mc_1\times\mc_2=\{U_1(m_1,j_1)\}_{m_1=1,j_1=1}^{\sm_1,\sj_1}\times\{U_2(m_2,j_2)\}_{m_2=1,j_2=1}^{\sm_2,\sj_2}$ be a random product codebook, in which the codebooks $\mc_1$ and $\mc_2$ are generated independently and the elements of the codebook $\mc_k,k=1,2$   are drawn independently from $q_{U_k}$. Thus the codebook is generated according to $r_{U_1U_2}=q_{U_1}q_{U_2}\neq q_{U_1U_2}$. Here $J_1,J_2$ are introducing redundancy but since it will be decoded at the receiver we can view these as  dummy messages.

\subsubsection*{Encoding} Instead of using conventional mutual covering, we use an SMC which acts as follows. Given $m_1,m_2$, the SMC chooses indices $j_1,j_2$ with the probability
\begin{align}
P_{\enc}(j_1,j_2|m_1,m_2)=\dfrac{2^{\im_q(U_1(m_1,j_1);U_2(m_2,j_2))}}{\sum_{\tilde{j}_1,\tilde{j}_2}2^{\im_q(U_1(m_1,\tilde{j}_1);U_2(m_2,\tilde{j}_2))}}.\label{eq:lastmn}
\end{align}
Then the encoder transmits $x(U_1(m_1,j_1),U_2(m_2,j_2))$ through the channel. Observe that we generate codewords according to $r_{U_1U_2}$ but compute the informations $\im_q$ using $q_{U_1U_2}$. This resembles the Marton coding scheme where we generate $U_1^n$ and $U_2^n$ independently but choose the jointly typical ones for transmission.
\subsubsection*{Decoding} We again use an SMC for decoding. Observing $y_k$, decoder $k$  uses the following SMC to find both the message $m_k$ and the dummy message $j_k$:
 \[
P_{\dec_k}(\hat{m}_k,\hat{j}_k|y)=\dfrac{2^{\im_q(y_k;U_k(\hat{m}_k,\hat{j}_k))}}{\sum_{\bar{m}_k,\bar{j}_k}2^{\im_q(y_k;U_k(\bar{m}_k,\bar{j}_k))}}.
\] 
\subsubsection*{Analysis} Observe that the joint distribution of random variables factors as,
\begin{align*}
P(m_{1:2},j_{1:2}&,y_{1:2},\hat{m}_{1:2},\hat{j}_{1:2})=\frac{1}{\sm_1\sm_2}P_{\enc}(j_{1:2}|m_{1:2})\\&q\big(y_{1:2}|U_1(m_1,j_1),U_2(m_2,j_2)\big)\prod_{k=1}^2P_{\dec_k}(\hat{m}_k,\hat{j}_k|y_k),
\end{align*}
 where $q\big(y_{1:2}|U_1(m_1,j_1),U_2(m_2,j_2)\big)$ is equal to $q_{Y_{1:2}|X}\big(y_{1:2}|x(U_1(m_1,j_1),U_2(m_2,j_2))\big)$. We make an error if either of the decoders fail.  The probability of correct decoding can be bounded from below by $\mathsf{P}[C]\ge\sum_{m_{1:2},j_{1:2},y_1}P(m_{1:2},j_{1:2},y_1,y_2,\hat{M}_{1:2}=m_{1:2}, \hat{J}_{1:2}=j_{1:2})$ , hence skipping similar symmetry arguments we have:
\begin{IEEEeqnarray}{rCl}
\e\mathsf{P}[C]&\ge&
                         \e\sum_{y_1,y_2}\sj_1\sj_2\dfrac{2^{\im_q(U_1(1,1);U_2(1,1))}}{\sum_{\tilde{j}_1,\tilde{j}_2}2^{\im_q(U_1(1,\tilde{j}_1);U_2(1,\tilde{j}_2))}}q(y_1|U_1(1,1),U_2(1,1))\prod_{k=1}^2\dfrac{2^{\im_q(y_k;U_k(1,1))}}{\sum_{\bar{m}_k,\bar{j}_k}2^{\im_q(y_k;U_k(\bar{m}_k,\bar{j}_k))}}\label{eq:201}\\
                         &\ge&\sum_{y_1,y_2}\textcolor{blue!90!black}{\e_{U_{1:2}(1,1)}}\left(\dfrac{\textcolor{blue!90!black}{\sj_1\sj_22^{\im_q(U_1(1,1);U_2(1,1))}}}{\textcolor{red!60!blue}{\e_{\mc|U_{1:2}(1,1)}\sum_{\tilde{j}_1,\tilde{j}_2}2^{\im_q(U_1(1,\tilde{j}_1);U_2(1,\tilde{j}_2))}}}\right.
\n
                         &&\left.\quad~~~~~~~~~~~~~~~~~~~~~~~~~~~~~~~~~~~~~~~ \textcolor{blue!90!black}{q(y_1|U_{1:2}(1,1))}\prod_{k=1}^2\dfrac{\textcolor{blue!90!black}{2^{\im_q(y_k;U_k(1,1))}}}{\textcolor{red!60!blue}{\e_{\mc|U_{1:2}(1,1)}\sum_{\bar{m}_k,\bar{j}_k}2^{\im_q(y_k;U_k(\bar{m}_k,\bar{j}_k))}}}\right)\IEEEeqnarraynumspace\label{eq:202}\\
                         &\ge&\sum_{y_1,y_2}\e_{U_{1:2}(1,1)}\left(\dfrac{\sj_1\sj_22^{\im_q(U_1(1,1);U_2(1,1))}}{2^{\im_q(U_1(1,1);U_2(1,1))}+\sj_1\sj_2}q(y_1|U_{1:2}(1,1))\prod_{k=1}^2\dfrac{2^{\im_q(y_k;U_k(1,1))}}{2^{\im_q(y_k;U_k(1,1))}+\sm_k\sj_k}\right)\label{eq:203}\\
                         &=&\sum_{u_1,u_2,y_1,y_2}q(u_1,u_2,y_1)\dfrac{\sj_1\sj_2}{2^{\im_q(u_1;u_2)}+\sj_1\sj_2}\prod_{k=1}^2\dfrac{2^{\im_q(y_k;u_k)}}{2^{\im_q(y_k;u_k)}+\sm_k\sj_k}\IEEEeqnarraynumspace\label{eq:204}\\
                         &=&\e_{q} \dfrac{1}{(1+(\sj_1\sj_2)^{-1}2^{\im_q(U_1;U_2)})\prod_{k=1}^2(1+\sm_k\sj_k2^{-\im_q(U_k;Y_k)})}\label{eq:205}
\end{IEEEeqnarray}
where \eqref{eq:201} is due to symmetry, the \underline{main step} \eqref{eq:202} follows from Jensen inequality for the 
 three-valued convex function $f(x_1,x_2,x_3)=\dfrac{1}{x_1x_2x_3}$ on $\mathbb{R}^3_+$.  The expectation in \eqref{eq:202} is over $U_{1:2}(1,1)$ of the codebook generation distributed according to $r_{U_1U_2}$. Equation \eqref{eq:203} follows from the following equations:

\begin{align}
 \e_{\mc|U_{1:2}(1,1)}&2^{\im_q(U_1(1,1);U_2(1,\tilde{j}_2))}=\sum_{u_2}q(u_2)2^{\im_q(U_1(1,1);u_2)}\n
&=\sum_{u_2}q(u_2|U(1,1))=1, \ \tilde{j}_2\neq1, \label{eq:203-1}\\
\e_{\mc|U_{1:2}(1,1)}&2^{\im_q(U_1(1,\tilde{j}_1);U_2(1,1))}=1,\ \tilde{j}_1\neq1,\label{eq:203-2}\\
\e_{\mc|U_{1:2}(1,1)}&2^{\im_q(U_1(1,\tilde{j}_1);U_2(1,\tilde{j}_2))}=1,\ \tilde{j}_1\neq1\ \tilde{j}_2\neq1,\label{eq:203-3}
\end{align}
where in \eqref{eq:203-1} we use the
 fact that $U_2(1,\tilde{j}_2)$ is independent of $(U_1(1,1),U_2(1,1))$ for $\bar{j}_2\neq 1$ and generated according to $q_{U_2}$. \eqref{eq:203-2} and \eqref{eq:203-3} follows similarly. Finally  and \eqref{eq:204} follows from the fact that $(U_1(1,1),U_2(1,1))$ is  generated according to product distribution $r_{U_1U_2}=q_{U_1}q_{U_2}$.
\end{IEEEproof}
\begin{remark}
Unlike the Gelfand-Pinsker problem, the SMC encoder used is \eqref{eq:lastmn} cannot be written in the form of a SLC encoder. The SLC encoder has the following form:
 
\begin{align*}
P_{\enc}^{SLC}(j_1,j_2|m_1,m_2)=\dfrac{q(U_1(m_1,j_1),U_2(m_2,j_2))}{\sum_{\tilde{j}_1,\tilde{j}_2}q(U_1(m_1,\tilde{j}_1),U_2(m_2,\tilde{j}_2))}.\label{eq:lastmn}
\end{align*}
We use SMC instead of SLC, because the analysis of SLC is challenging. 
\end{remark}
\subsubsection{Second Order achievability of broadcast channel}
\begin{theorem}\label{thm:GFB}
Given a memoryless broadcast channel $q_{Y_1,Y_2|X}$,  for any $(q_{U_1U_2},x(u_1,u_2))$, the pair $(R_1,R_2)$ is $(n,\epsilon)$-achievable, if there exists $(\tR_1,\tR_2)$ such that
\begin{align}
\left[\begin{array}{c}~~\tR_1+\tR_2\\-R_1-\tR_1\\-R_2-\tR_2\end{array}\right]\in\mathbb{I}_{\mathsf{BC}}+\dfrac{1}{\sqrt{n}}\mq^{-1}(\mathbb{V}_{\mathsf{BC}},\ep)+O\left(\dfrac{\log n}{n}\right)
\end{align}
where
\be
\mathbb{I}_{\mathsf{BC}}=\e_{U_1U_2Y_1Y_2}I_{\mathsf{BC}},~~~~\mathbb{V}_{\mathsf{BC}}=\mathsf{Cov}(I_{\mathsf{BC}}),
\end{equation}
in which
\begin{equation}
I_{\mathsf{BC}}=\left[\begin{array}{c} \im(U_1;U_2)\\ -\im(U_1;Y_1)\\-\im(U_2;Y_2)\end{array}\right].
\end{equation}

\end{theorem}
\begin{IEEEproof}
The proof uses \eqref{eq:200} in a way similar to the proof of Theorem \ref{thm:GPFB} and hence is omitted. 
\end{IEEEproof}

\section{Multi-terminal lossy source coding problems}
\label{MTLSC}
To illustrate the application of our technique to multi-terminal lossy source coding problems, we study the problems of Berger-Tung, Heegard-Berger/Kaspi and Multiple-description in this section.
Throughout this section we use $s$ for source and $\hat{s}$ for its reconstruction. Since $\hat{s}$ is created from a set of rv's available at the decoder, we follow El Gamal and Kim's notation \cite{NIT} to also use $\hat{s}$ as a function of the rv's available at the decoder. For instance if the decoder has rv $Y$, we use a decoding function $\hat{s}(y)$.

\subsection{Berger-Tung}

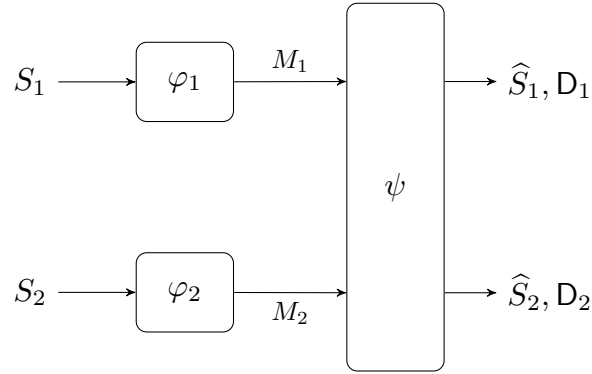
\begin{figure}
 \begin{center}
  \begin{tikzpicture}[scale=1.4,>=
   stealth'
  ]
\def\bscale{1.5}
\tikzstyle{enc}=[scale=.8,draw=black, text width=2.4em, rounded corners,
    text centered, minimum height=2.5em
    ]
    \tikzstyle{ch}=[scale=.8,draw=black, minimum width=2.6em, rounded corners,
    text centered, minimum height=6em,draw=blue,
    ]
\tikzstyle{ann} = [above, text width=2em]
\tikzstyle{dec} = [enc, text width=2.4em, 
    minimum height=2em,fill=gray!70!white,draw=black]
 \node (d0) [coordinate]{};
 \path (d0.north)+(0,1) node (d1) [enc,scale=\bscale]{$\varphi_1$};
  \path (d0.south)+(0,-1) node (d2) [enc,scale=\bscale]{$\varphi_2$}; 
   \path (d0.east)+(2,0) node (e0) [enc,scale=\bscale,minimum height=11.6em]{$\psi$}; 
  \path (e0.65)+(1,0) node (e01) [scale=1.2]{$\widehat{S}_1,\mathsf{D}_1$};
    \path (e0.-65)+(1,0) node (e010) [scale=1.2]{$\widehat{S}_2,\mathsf{D}_2$};
    \path (d1.west)+(-1,0) node (d10) [scale=1.2]{${S}_1$};
     \path (d2.west)+(-1,0) node (d20) [scale=1.2]{${S}_2$};
  \path[draw,<-] (e0.115)-- node(m)[above]{$M_1$}(d1);
    \path[draw,<-] (e0.-115)-- node(m2)[below]{$M_2$}(d2);
         \draw[<-](e01)--(e0.65); 
                  \draw[<-](e010)--(e0.-65); 
          \draw[<-](d1)--(d10);
           \draw[<-](d2)--(d20);
  \end{tikzpicture} 
  \caption{Distributed lossy compression system}\label{fig:BT}
  \end{center}\end{figure}

 Consider the problem of distributed lossy compression depicted in Fig. \ref{fig:BT}. Let $q_{S_1S_2}$ be the joint distribution of the sources and $\mathsf{d}_k(s_k,\hat{s}_k), k=1,2,$ be two distortion measures. We  will use the \emph{probability of excess distortion} as the criterion for measuring the reliability of the system. 
 
 \begin{definition}
 An $(\sm_1,\sm_2)$-code for distributed lossy compression consists of (possibly stochastic) encoders $\varphi_k:\ms_k\mapsto[1:\sm_k], k=1,2$, and  (possibly stochastic) decoder $\psi:[1:\sm_1]\times[1:\sm_2]\mapsto\widehat{S}_1\times \widehat{S}_2$. Given a pair of distortion levels $(\mathsf{D}_1,\mathsf{D}_2)$, the probability of excess distortion is defined by,
\[\mathsf{P}[\me;\mathsf{D}_1,\mathsf{D}_2]:=\mathsf{P}[\mathsf{d}_1(S_1,\widehat{S}_1)>\mathsf{D}_1\ \vee\ \mathsf{d}_2(S_2,\widehat{S}_2)>\mathsf{D}_2].\] 
Also, we define the probability of \emph{no-excess distortion} by 
\[\mathsf{P}[C;\mathsf{D}_1,\mathsf{D}_2]:=\mathsf{P}[\mathsf{d}_1(S_1,\widehat{S}_1)\leq \mathsf{D}_1,\  \mathsf{d}_2(S_2,\widehat{S}_2)\leq\mathsf{D}_2].\] 
 \end{definition}
 
 We prove a one-shot version of the result of Berger-Tung \cite{berger,tung} for this problem. 
 \begin{theorem}
 Given any pmf $q_{U_1|S_1}q_{U_2|S_2}$ and functions $\hat{s}_k(u_1,u_2),k=1,2$, there is an  $(\sm_1,\sm_2)$-code for a single use of the sources whose probability of \emph{no-excess distortion} is bounded from below by
 \be
\mathsf{P}[C;\mathsf{D}_1,\mathsf{D}_2]\ge \e\dfrac{\mathbf{1}\left\{\mathsf{d}_1\left(S_1,\hat{s}_1(U_1,U_2)\right)\le \mathsf{D}_1,\mathsf{d}_2\left(S_2,\hat{s}_2(U_1,U_2)\right)\le \mathsf{D}_2\right\}}{(1+\sj_1^{-1}2^{\im(S_1;U_1)})
                        (1+\sj_2^{-1}2^{\im(S_2;U_2)})\left(1+(\sj_2\sm_2^{-1}+\sj_1\sm_1^{-1}+\sj_1\sj_2(\sm_1\sm_2)^{-1})2^{-\im(U_1;U_2)}\right)},
 \ee
  where $\sj_k\ge\sm_k, k=1,2$ are arbitrary integers.
  Moreover, loosening this bound gives the following upper bound on the probability of excess distortion 
of the code,
  \begin{align}
 \mathsf{P}&\left[\mathsf{d}_1\left(S_1,\hat{s}_1(U_1,U_2)\right)> \mathsf{D}_1, \mathsf{or}
 ~\mathsf{d}_2\left(S_2,\hat{s}_2(U_1,U_2)\right)> \mathsf{D}_2
 , \mathsf{or}\ \right. \n
 &~~~\log\sj_1-\im(S_1;U_1)<\gamma, \mathsf{or}\ ~ \log\sj_2-\im(S_2;U_2)<\gamma, \mathsf{or}\ ~ \im(U_1;U_2)-\log\dfrac{\sj_1\sj_2}{\sm_1\sm_2}<\gamma\Big]+15\times2^{-\gamma}.
\end{align}
  \end{theorem}
\begin{remark}
 The term $1+\sj_k^{-1}2^{\im(S_k;U_k)}, ~~k=1,2$ corresponds to a covering of $S_k$ with $U_k$ in the asymptotic regime. The alphabet size of $U_k$ is $\sj_k$. We use a random binning of $U_k$, mapping it from a set of size $\sj_k$ to a set of size $\sm_k$. This explains the inequality $\sj_k\ge\sm_k$ for $k=1,2$. The term $\left(1+(\sj_2\sm_2^{-1}+\sj_1\sm_1^{-1}+\sj_1\sj_2(\sm_1\sm_2)^{-1})2^{-\im(U_1;U_2)}\right)$ corresponds to a mutual packing lemma in the asymptotic regime.
\end{remark} 

\begin{IEEEproof} 
 We only prove the lower bound on probability of no-excess distortion. Derivation of the loosened bound is similar to that of Gelfand-Pinsker and is thus omitted. 
 \subsubsection*{Random codebook generation}Let $\mc=\mc_1\times\mc_2=\{U_1(j_1)\}_{j_1=1}^{\sj_1}\times\{U_2(j_2)\}_{j_2=1}^{\sj_2}$ be a random product codebook, in which the codebooks $\mc_1$ and $\mc_2$ are generated independently and the elements of the codebook $\mc_k,k=1,2$   are drawn independently from $q_{U_k}$. Thus the codebook is generated according to $r_{U_1U_2}=q_{U_1}q_{U_2}\neq q_{U_1U_2}$. 
 \subsubsection*{Random binning} Let $\mb_k:[1:\sj_k]\mapsto[1:\sm_k], k=1,2$ be two independent random mappings (binning), in which $\mb_k$ maps each element of $[1:\sj_k]$ uniformly and independently to the set $[1:\sm_k]$.   
\subsubsection*{Encoding} Encoder $k=1,2$  uses an SMC  followed by a random binning to obtain the index $m_k$. First  given $s_k$, the SMC chooses an index $j_k$ with the probability
\[
P_{\enc_k}(j_k|s_k)=\dfrac{2^{\im_q(s_k;U(j_k))}}{\sum_{\tilde{j}_k}2^{\im_q(s_k;U(\tilde{j}_k))}}.
\]  
Then the encoder transmits $m_k=\mb_k(j_k)$ to the decoder. 
\subsubsection*{Decoding} We use a modified SMC for decoding w.r.t. a receiving indices $(m_1,m_2)$. Observing $(m_1,m_2)$, decoder   uses the following modified SMC to find both the $j_1$ and $j_2$ and thus $(U_1,U_2)$:
 \[
P_{\dec}(\hat{j}_1,\hat{j}_2|m_1,m_2)=\dfrac{2^{\im_q(U_1(\hat{j}_1);U_2(\hat{j}_2))}\mathbf{1}\{B_1(\hat{j}_1)=m_1,B_2(\hat{j}_2)=m_2\}}{\sum_{\bar{j}_1,\bar{j}_2}2^{\im_q(U_1(\bar{j}_1);U_2(\bar{j}_2))}\mathbf{1}\{B_1(\bar{j}_1)=m_1,B_2(\bar{j}_2)=m_2\}}.
\]

\textbf{Remark:} Observe that the above SMC resembles a joint-typical decoder for the mutual packing lemma in the asymptotic regime. It can be considered as a dual of the SMC encoder of equation \eqref{eq:lastmn} that corresponded to a mutual covering lemma in the asymptotic regime. 

 Given $m_1$ and $m_2$, the decoder chooses among the pairs $(U_1(j_1),U_2(j_2))$ assigned to the bin $(m_1,m_2)$. The higher the information between $U_1(j_1)$ and $U_2(j_2)$, the more likely we choose it at the decoder. 
Finally, decoder computes $\hat{s}_k(U_1(\hat{j_1}),U_2(\hat{j}_2))$ as the estimate of $S_k$.
\subsubsection*{Analysis} Observe that the joint distribution of random variables factors as,
\begin{align*}
P(s_{1:2},m_{1:2},j_{1:2},\hat{j}_{1:2})=q(s_1,s_2)\prod_{k=1}^2P_{\enc_k}(j_{k}|s_k)\mathbf{1}\{B_1({j}_1)=m_1,B_2({j}_2)=m_2\}P_{\dec}(\hat{j}_1,\hat{j}_2|m_1,m_2).
\end{align*}
We consider two error events:
 \begin{enumerate}
 \item The decoder fails to recover the correct pair $(j_1,j_2)$, i.e. $(\hat{j}_1,\hat{j}_2)\neq (j_1, j_2)$.
 \item One of the distortions corresponding to the pair $(j_1,j_2)$ exceeds the designated levels, i.e. 
 $\mathsf{d}_k\big(s_k, \hat{s}_k(U_1({j_1}),U_2({j}_2))\big)>\mathsf{D}_k$ for some $k\in \{1,2\}$.
 \end{enumerate}
  The probability of correct decoding can be bounded from below by 
\begin{align*}\mathsf{P}[C]\ge\sum_{s_{1:2},m_{1:2},j_{1:2}}&P(s_{1:2},m_{1:2},j_{1:2},\hat{J}_1=j_1, \hat{J}_2=j_2)\\&~~~~~~~~\times\mathbf{1}\left\{\mathsf{d}_1\left(s_1,\hat{s}_1(U_1(j_1),U_2(j_2))\right)\le \mathsf{D}_1,\mathsf{d}_2\left(s_2,\hat{s}_2(U_1(j_1),U_2(j_2))\right)\le \mathsf{D}_2\right\},
\end{align*} 
 hence skipping similar symmetry arguments we have:
 \begin{IEEEeqnarray}{rCl}
\e\mathsf{P}&[C]&\ge
                         \e_{\mc,\mb}\sum_{s_1,s_2}\sm_1\sm_2\sj_1\sj_2q(s_1,s_2)\dfrac{2^{\im(s_1;U_1(1))}}{\sum_{\tilde{j}_1}2^{\im(s_1;U_1(\tilde{j}_1))}}\mathbf{1}(B_1(1)=1)\dfrac{2^{\im(s_2;U_2(1))}}{\sum_{\tilde{j}_2}2^{\im(s_2;U_2(\tilde{j}_2))}}\mathbf{1}(B_2(1)=1)\n&&~~~~~~\times\dfrac{2^{\im(U_1(1);U_2(1))}\mathbf{1}(B_1(1)=1,B_2(1)=1)}{\sum_{\bar{j}_1,\bar{j}_2}2^{\im(U_1(\bar{j}_1);U_2(\bar{j}_2))}\mathbf{1}\{B_1(\bar{j}_1)=1,B_2(\bar{j}_2)=1\}}\n&&~~~~~~\times\mathbf{1}\left\{\mathsf{d}_1\left(s_1,\hat{s}_1(U_1(1),U_2(1))\right)\le \mathsf{D}_1,\mathsf{d}_2\left(s_2,\hat{s}_2(U_1(1),U_2(1))\right)\le \mathsf{D}_2\right\}\label{eq:BT201}\\
                         &\ge&\sm_1\sm_2\sum_{s_1,s_2}q(s_1,s_2)\textcolor{blue!90!black}{\e_{U_{1:2}(1),B_{1:2}(1)}}\left(\dfrac{\textcolor{blue!90!black}{\sj_12^{\im(s_1;U_1(1))}}}{\textcolor{red!60!blue}{\e_{\mc|U_{1:2}(1)}\sum_{\tilde{j}_1}2^{\im(s_1;U_1(\tilde{j}_1))}}}\dfrac{\textcolor{blue!90!black}{\sj_22^{\im(s_2;U_2(1))}}}{\textcolor{red!60!blue}{\e_{\mc|U_{1:2}(1)}\sum_{\tilde{j}_2}2^{\im(s_2;U_2(\tilde{j}_2))}}}\right.\n&&~~~~~~\times\dfrac{\textcolor{blue!90!black}{2^{\im(U_1(1);U_2(1))}\mathbf{1}(B_1(1)=1,B_2(1)=1)}}{\textcolor{red!60!blue}{\e_{\mc,\mb|U_{1:2}(1),B_{1:2}(1)}\sum_{\bar{j}_1,\bar{j}_2}2^{\im(U_1(\bar{j}_1);U_2(\bar{j}_2))}\mathbf{1}\{B_1(\bar{j}_1)=1,B_2(\bar{j}_2)=1\}}}\n&&~~~~~~\times\textcolor{blue!90!black}{\mathbf{1}\left\{\mathsf{d}_1\left(s_1,\hat{s}_1(U_1(1),U_2(1))\right)\le \mathsf{D}_1,\mathsf{d}_2\left(s_2,\hat{s}_2(U_1(1),U_2(1))\right)\le \mathsf{D}_2\right\}}\Bigg)\label{eq:BT202}\\
                         &\ge&\sm_1\sm_2\sum_{s_1,s_2}q(s_1,s_2)\e_{U_{1:2}(1),B_{1:2}(1)}\left(\dfrac{2^{\im(s_1;U_1(1))}}{1+\sj_1^{-1}2^{\im(s_1;U_1(1))}}
                        \dfrac{2^{\im(s_2;U_2(1))}}{1+\sj_2^{-1}2^{\im(s_2;U_2(1))}}\right.\n&&~\times\dfrac{2^{\im(U_1(1);U_2(1))}\mathbf{1}(B_1(1)=1,B_2(1)=1)}{2^{\im(U_1(1);U_2(1))}\mathbf{1}(B_1(1)=1,B_2(1)=1)+\sj_2\sm_2^{-1}\mathbf{1}(B_1(1)=1)+\sj_1\sm_1^{-1}\mathbf{1}(B_2(1)=1)+\sj_1\sj_2(\sm_1\sm_2)^{-1}}\n&&~~~~~~\times\mathbf{1}\left\{\mathsf{d}_1\left(s_1,\hat{s}_1(U_1(1),U_2(1))\right)\le \mathsf{D}_1,\mathsf{d}_2\left(s_2,\hat{s}_2(U_1(1),U_2(1))\right)\le \mathsf{D}_2\right\}\Bigg)\label{eq:BT203}\\
                         &=&\sum_{s_1,s_2,u_1,u_2}q(s_1,s_2)q(u_1|s_1)q(u_2|s_2)\dfrac{1}{1+\sj_1^{-1}2^{\im(s_1;u_1)}}
                        \dfrac{1}{1+\sj_2^{-1}2^{\im(s_2;u_2)}}\n&&~~\qquad~~~\quad~~~\times\dfrac{\mathbf{1}\left\{\mathsf{d}_1\left(s_1,\hat{s}_1(u_1,u_2)\right)\le \mathsf{D}_1,\mathsf{d}_2\left(s_2,\hat{s}_2(u_1,u_2)\right)\le \mathsf{D}_2\right\}}{1+(\sj_2\sm_2^{-1}+\sj_1\sm_1^{-1}+\sj_1\sj_2(\sm_1\sm_2)^{-1})2^{-\im(u_1;u_2)}}\\
                        &=&\e_{q_{S_{1:2}U_{1:2}}}\dfrac{\mathbf{1}\left\{\mathsf{d}_1\left(S_1,\hat{s}_1(U_1,U_2)\right)\le \mathsf{D}_1,\mathsf{d}_2\left(S_2,\hat{s}_2(U_1,U_2)\right)\le \mathsf{D}_2\right\}}{(1+\sj_1^{-1}2^{\im(S_1;U_1)})
                        (1+\sj_2^{-1}2^{\im(S_2;U_2)})\left(1+(\sj_2\sm_2^{-1}+\sj_1\sm_1^{-1}+\sj_1\sj_2(\sm_1\sm_2)^{-1})2^{-\im(U_1;U_2)}\right)}\label{eq:BT204}
                         \end{IEEEeqnarray}
where \eqref{eq:BT201} is due to symmetry, the \underline{main step} \eqref{eq:BT202} follows from Jensen inequality for the 
 three-valued convex function $f(x_1,x_2,x_3)=\dfrac{1}{x_1x_2x_3}$ on $\mathbb{R}^3_+$.  The expectation in \eqref{eq:BT202} is over $U_{1:2}(1,1)$ of the codebook generation distributed according to $r_{U_1U_2}$. Equation \eqref{eq:BT203} follows from separating the cases $(\bar{j}_1=1, \bar{j}_2\neq 1)$, $(\bar{j}_1\neq 1, \bar{j}_2= 1)$, $(\bar{j}_1\neq 1, \bar{j}_2\neq 1)$ and $(\bar{j}_1=1, \bar{j}_2=1)$. This is discussed below. Lastly, equation \eqref{eq:BT204} follows from the facts that random codebook and random binning are independent and  $(U_1(1),U_2(1))$ is  generated according to product distribution $r_{U_1U_2}=q_{U_1}q_{U_2}$. 
 
\subsubsection*{Derivation of equation \eqref{eq:BT203}}
 In the following we consider each of the four case mentioned above separately. 
\begin{itemize}
\item Case 1, $(\bar{j}_1=1, \bar{j}_2\neq 1)$:
\begin{align}
 \e_{\mc,\mb|U_{1:2}(1),B_{1:2}(1)}&2^{\im(U_1(1);U_2(\bar{j}_2))}\mathbf{1}\{B_1(1)=1,B_2(\bar{j}_2)=1\}=\sum_{u_2}q(u_2)2^{\im(U_1(1);u_2)}\mathbf{1}\{B_1(1)=1\}\sm_2^{-1}\n
&=\sm_2^{-1}\sum_{u_2}q(u_2|U_1(1))\mathbf{1}\{B_1(1)=1\}=\sm_2^{-1}\mathbf{1}\{B_1(1)=1\}\label{eq:BT203-1}
\end{align}
where in \eqref{eq:BT203-1} we use the 
following facts:
\begin{enumerate}
\item $U_2(\bar{j}_2)$ is independent of $(U_1(1),U_2(1))$ for $\bar{j}_2\neq 1$ and generated according to $q_{U_2}$.
\item The random binning and random codebook are independent and $B_2(\bar{j}_2)$ has a uniform distribution over $[1:\sm_2]$. 
\end{enumerate}   

\item Case 2, $(\bar{j}_1\neq 1, \bar{j}_2=1)$: This is similar to the previous case and gives us …
\begin{align}
\e_{\mc,\mb|U_{1:2}(1),B_{1:2}(1)}&2^{\im(U_1(\bar{j}_1);U_2(1))}\mathbf{1}\{B_1(\bar{j}_1)=1,B_2(1)=1\}=\sm_1^{-1}\mathbf{1}\{B_2(1)=1\}\label{eq:BT203-2}
\end{align}

\item Case 3: $(\bar{j}_1\neq 1, \bar{j}_2\neq 1)$: In this case $U_1(\bar{j}_1), U_2(\bar{j}_2)$ are independent of $U_{1:2}(1),B_{1:2}(1)$, and are generated according to $q_{U_1}q_{U_2}$.
\begin{align}
 \e_{\mc,\mb|U_{1:2}(1),B_{1:2}(1)}&2^{\im(U_1(\bar{j}_1);U_2(\bar{j}_2))}\mathbf{1}\{B_1(\bar{j}_1)=1,B_2(\bar{j}_2)=1\}=\sum_{u_1,u_2}q(u_1)q(u_2)2^{\im(u_1;u_2)}(\sm_1\sm_2)^{-1}\n
&=(\sm_1\sm_2)^{-1}\sum_{u_1,u_2}q(u_1,u_2)=(\sm_1\sm_2)^{-1}.\label{eq:BT203-3}
\end{align}

\item Case 4: $(\bar{j}_1= 1, \bar{j}_2=1)$: 
\begin{align}
\e_{\mc,\mb|U_{1:2}(1),B_{1:2}(1)}&2^{\im(U_1(1);U_2(1))}\mathbf{1}\{B_1(1)=1,B_2(1)=1\}=2^{\im(U_1(1);U_2(1))}\mathbf{1}\{B_1(1)=1,B_2(1)=1\}.
\end{align}

\end{itemize}
\end{IEEEproof}
\subsection{Heegard-Berger/Kaspi}
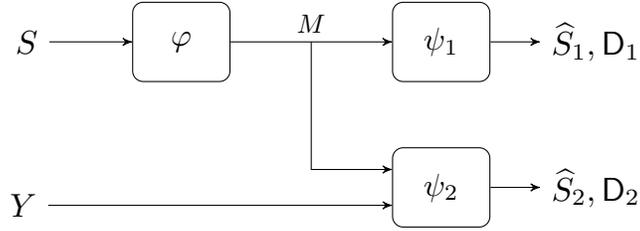
\begin{figure}
\begin{center}\begin{tikzpicture}[scale=1.4,>=stealth']
\def\bscale{1.5}
\tikzstyle{enc}=[scale=.8,draw=black, text width=2.4em, rounded corners,
    text centered, minimum height=2.5em
    ]
    \tikzstyle{ch}=[scale=.8,draw=black, minimum width=2.6em, rounded corners,
    text centered, minimum height=6em,draw=blue,
    ]
\tikzstyle{ann} = [above, text width=2em]
\tikzstyle{dec} = [enc, text width=2.4em, 
    minimum height=2em,fill=gray!70!white,draw=black]
 \node (d0) [enc,scale=\bscale]{$\psi_2$};
 \path (d0.north)+(0,1) node (d1) [enc,scale=\bscale]{$\psi_1$};
       \path (d1.west)+(-2,0) node (e1) [enc,scale=\bscale]{$\varphi$}; 
          \path (d0.east)+(1,0) node (d00) [scale=1.25]{$\widehat{S}_2,\mathsf{D}_2$};
    \path (d1.east)+(1,0) node (d10) [scale=1.25]{$\widehat{S}_1,\mathsf{D}_1$};
  \path (e1.west)+(-1,0) node (e0) [scale=1.25]{$S$};
  \path (d0.200)+(-3.5,0) node (e00) [scale=1.25]{$Y$};
  \draw[->](e0)--(e1);  
    \draw[->](e00)--(d0.200);  
   \path [draw,->] (e1)--node(m)[above]{\textcolor{black}{$M$}} (d1);
   \draw[->](m)|-(d0.160); 
                \draw[->](d0)--(d00);           \draw[->](d1)--(d10);
  \end{tikzpicture}
  \caption{Lossy source coding when side information is absent.}\label{fig:HB-K}
  \end{center}

  \end{figure} 
Consider the problem of  lossy source coding when side information may be absent, depicted in Fig. \ref{fig:HB-K}. Let $q_{S}$ be the  distribution of the source and $\mathsf{d}_k(s,\hat{s}_k), k=1,2,$ be two distortion measures. We  will use the \emph{probability of excess distortion} as the criterion for measuring the reliability of the system.

 \begin{definition}
 An $\sm$-code for distributed lossy compression consists of (possibly stochastic) encoder $\varphi:\ms\mapsto[1:\sm]$, and  (possibly stochastic) decoders $\psi_1:[1:\sm]\mapsto\widehat{S}_1$ and $\psi_2:[1:\sm]\times\my\mapsto\widehat{S}_2$. Given a pair of distortion levels $(\mathsf{D}_1,\mathsf{D}_2)$, the probability of excess distortion is defined by,
\[\mathsf{P}[\me;\mathsf{D}_1,\mathsf{D}_2]:=\mathsf{P}[\mathsf{d}_1(S,\widehat{S}_1)>\mathsf{D}_1\ \vee\ \mathsf{d}_2(S,\widehat{S}_2)>\mathsf{D}_2].\] 
Also, we define the probability of \emph{no-excess distortion} by 
\[\mathsf{P}[C;\mathsf{D}_1,\mathsf{D}_2]:=\mathsf{P}[\mathsf{d}_1(S,\widehat{S}_1)\leq \mathsf{D}_1,\  \mathsf{d}_2(S,\widehat{S}_2)\leq \mathsf{D}_2].\] 
 \end{definition}
 \begin{remark} Setting $\mathsf{D}_1=\infty$, this problem reduces to the Wyner-Ziv problem. A one-shot result for the Wyner-Ziv problem has been concurrently obtained by \cite{watanabe}.
 \end{remark}
 We prove a one-shot version of the result of Heegard-Berger/Kaspi \cite{heegard,kaspi} for this problem. 
 
  \begin{theorem}
 Given any  $q_{WU|S}$ and functions $\hat{s}_1(w),\hat{s}_2(w,u,y)$, there is an  $\sm$-code for a single use of the source whose probability of \emph{no-excess distortion} is bounded from below by
 \be
\mathsf{P}[C;\mathsf{D}_1,\mathsf{D}_2]\ge \e\dfrac{{\mathbf{1}\left\{\mathsf{d}_1\left(S,\hat{s}_1(W)\right)\le \mathsf{D}_1,\mathsf{d}_2\left(S,\hat{s}_2(W,U,Y)\right)\le \mathsf{D}_2\right\}}}{\left({1+\sm_1^{-1}2^{\im(S;W)}+(\sm_1\sj_2)^{-1}2^{\im(S;W,U)}}\right)
                        \left({1+\sj_2\sm_2^{-1}2^{-\im(Y;U|W)}}\right)},
 \ee
  where $\sm_1,\sm_2,\sj_2$ are  integers such that $\sm=\sm_1\sm_2$ and $\sj_2\ge\sm_2$.
  Moreover, loosening this bound gives the following upper bound on the probability of excess distortion 
of the code,
 \begin{align}
 \mathsf{P}&\left[\mathsf{d}_1\left(S,\hat{s}_1(W)\right)> \mathsf{D}_1, \mathsf{or}
 ~\mathsf{d}_2\left(S,\hat{s}_2(W,U,Y)\right)> \mathsf{D}_2
 , \mathsf{or}\ \right. \n
 &~~~\log\sm_1-\im(S;W)<\gamma, \mathsf{or}\ ~ \log\sm_1\sj_2-\im(S;WU)<\gamma, \mathsf{or}\ ~ \im(Y;U|W)-\log\sj_2\sm_2^{-1}<\gamma\Big]+5\times2^{-\gamma}.
\end{align}
  \end{theorem}

\begin{IEEEproof} 
 We only prove the lower bound on probability of no-excess distortion. Derivation of the loosened bound is similar to that of Gelfand-Pinsker and is thus omitted. 
 \subsubsection*{Random codebook generation}
 \begin{itemize}
 \item $\mc_1=\{W(m_1)\}_{m_1=1}^{\sm_1}$ is a random codebook whose elements are drawn independently from $q_W$.
 \item For each $m_1\in[1:\sm_1]$, let $\mc_2(m_1)=\{U(m_1,j_2)\}_{j_2=1}^{\sj_2}$ be a random codebook whose elements are drawn independently from $q_{U|W}(.|W(m_1))$. Moreover the codebooks $\mc_2(m_1),m_1\in[1:\sm_1]$ are generated independently.
 \item Let $\mc$ be the set of all codewords. 
 \end{itemize}
 
 \subsubsection*{Random binning} Let $\mb:[1:\sj_2]\mapsto[1:\sm_2]$ be a random mapping (binning), where $\mb$ maps each element of $[1:\sj_2]$ uniformly and independently to the set $[1:\sm_2]$.   
\subsubsection*{Encoding} Encoder  uses an SMC  followed by a random binning to obtain the pair $(m_1,m_2)$. First  given $s$, the SMC chooses a pair $(m_1,j_2)$ with the probability
\[
P_{\enc}(m_1,j_2|s)=\dfrac{2^{\im_q(s;W(m_1),U(m_1,j_2))}}{\sum_{\tilde{m}_1,\tilde{j}_2}2^{\im_q(s;W(\tilde{m}_1),U(\tilde{m}_1,\tilde{j}_2))}}.
\]  
Then the encoder computes $m_2=\mb(j_2)$ and sends $m=(m_1,m_2)$ to the decoders. 
\subsubsection*{Decoding} Decoder 1 outputs $\hat{s}_1(W(m_1))$ as the estimate of $S$. Decoder 2 uses a modified SMC for decoding w.r.t. a received index $(m_1,m_2)$. Observing $(m_1,m_2)$, decoder 2  uses the following modified SMC to find $j_2$ and thus $(W(m_1),U(m_1,j_2))$:
 \[
P_{\dec}(\hat{j}_2|m_1,m_2)=\dfrac{2^{\im_q(y;U(m_1,\hat{j}_2)|W(m_1))}\mathbf{1}(B(\hat{j}_2)=m_2)}{\sum_{\bar{j}_2}2^{\im_q(y;U(m_1,\bar{j}_2)|W(m_1))}\mathbf{1}(B(\bar{j}_2)=m_2)}.
\]
Observe that the above SMC resembles a joint-typical decoder of the asymptotic regime. 
Finally, decoder 2 computes $\hat{s}_2(W(m_1),U(m_1,\hat{j}_2),y)$ as the estimate of $S$.
\subsubsection*{Analysis} Observe that the joint distribution of random variables factors as,
\begin{align*}
P(s,y,m_{1:2},j_{2},\hat{j}_{2})=q(s,y)P_{\enc}(m_1,j_{2}|s)\mathbf{1}\{B(\hat{j}_2)=m_2\}P_{\dec}(\hat{j}_2|m_1,m_2).
\end{align*}

We consider two error events:
 \begin{enumerate}
 \item Decoder 2 fails to recover the correct $j_2$, i.e. $\hat{j}_2\neq j_2$.
 \item The distortion of decoder 1, or that of decoder 2 corresponding to $j_2$ exceeds the designated levels, i.e. 
 $\mathsf{d}_1\big(s, \hat{s}_1(W(m_1)))\big)>\mathsf{D}_1$ or
$\mathsf{d}_2\big(s, \hat{s}_2(W(m_1),U(m_1,\hat{j}_2),y)\big)>\mathsf{D}_2$.
 \end{enumerate}

  The probability of correct decoding can be bounded from below by 
\begin{align*}\mathsf{P}[C]\ge\sum_{s,y,m_{1:2},j_{2}}&P(s,y,m_{1:2},j_{2},\hat{J}_2=j_2)\\&~~~~~~~~\times\mathbf{1}\left\{\mathsf{d}_1\left(s,\hat{s}_1(W(m_1))\right)\le \mathsf{D}_1,\mathsf{d}_2\left(s,\hat{s}_2(W(m_1),U(m_1,j_2),y)\right)\le \mathsf{D}_2\right\},
\end{align*} 
hence skipping similar symmetry arguments we have:
\begin{IEEEeqnarray}{rCl}
\e\mathsf{P}&[C]&\ge
                         \e_{\mc,\mb}\sum_{s,y}\sm_1\sm_2\sj_2q(s,y)\dfrac{2^{\im(s;W(1),U(1,1))}}{\sum_{\tilde{m}_1,\tilde{j}_2}2^{\im(s;W(\tilde{m}_1),U(\tilde{m}_1,\tilde{j}_2))}}\mathbf{1}(B(1)=1)\dfrac{2^{\im(y;U(1,1)|W(1))}\mathbf{1}(B(1)=1)}{\sum_{\bar{j}_2}2^{\im(y;U(1,\bar{j}_2)|W(1))}\mathbf{1}(B(\bar{j}_2)=1)}\n&&~~~~~~\times\mathbf{1}\left\{\mathsf{d}_1\left(s,\hat{s}_1(W(1))\right)\le \mathsf{D}_1,\mathsf{d}_2\left(s,\hat{s}_2(W(1),U(1,1),y)\right)\le \mathsf{D}_2\right\}\label{eq:HB201}\\
                         &\ge&\sm_2\sum_{s,y}q(s,y)\textcolor{blue!90!black}{\e_{W(1),U(1,1),B(1)}}\left(\dfrac{\textcolor{blue!90!black}{\sm_1\sj_2 2^{\im(s;W(1),U(1,1))}}}{\textcolor{red!60!blue}{\e_{\mc|W(1),U(1,1)}\sum_{\tilde{m}_1,\tilde{j}_2}2^{\im(s;W(\tilde{m}_1),U(\tilde{m}_1,\tilde{j}_2))}}}\right.\n&&~~~~~~\qquad\qquad\qquad\qquad~~~~~~~~~\times\dfrac{\textcolor{blue!90!black}{2^{\im(y;U(1,1)|W(1))}\mathbf{1}(B(1)=1)}}{\textcolor{red!60!blue}{\e_{\mc,\mb|W(1),U(1,1),B(1)}\sum_{\bar{j}_2}2^{\im(y;U(1,\bar{j}_2)|W(1))}\mathbf{1}(B(\bar{j}_2)=1)}}\n
                         &&~~~~~~\qquad\qquad\qquad\qquad\times\textcolor{blue!90!black}{\mathbf{1}\left\{\mathsf{d}_1\left(s,\hat{s}_1(W(1))\right)\le \mathsf{D}_1,\mathsf{d}_2\left(s,\hat{s}_2(W(1),U(1,1),y)\right)\le \mathsf{D}_2\right\}}\Bigg)\label{eq:HB202}\\
                         &\ge&\sm_2\sum_{s,y}q(s,y){\e_{W(1),U(1,1),B(1)}}\left(\dfrac{{\sm_1\sj_2 2^{\im(s;W(1),U(1,1))}}}{\sm_1\sj_2+\sj_22^{\im(s;W(1))}+2^{\im(s;W(1),U(1,1))}}\right.\n&&~~~~~~\qquad\qquad\qquad\qquad~~~~~~~~~\times\dfrac{{2^{\im(y;U(1,1)|W(1))}\mathbf{1}(B(1)=1)}}{{2^{\im(y;U(1,1)|W(1))}\mathbf{1}(B(1)=1)}+\sj_2\sm_2^{-1}}\n
                         &&~~~~~~\qquad\qquad\qquad\qquad\times{\mathbf{1}\left\{\mathsf{d}_1\left(s,\hat{s}_1(W(1))\right)\le \mathsf{D}_1,\mathsf{d}_2\left(s,\hat{s}_2(W(1),U(1,1),y)\right)\le \mathsf{D}_2\right\}}\Bigg)\label{eq:HB203}\\                       
                         &=&\sum_{s,y,w,u}q(s,y)q(w,u|s)\left(\frac{1}{1+\sm_1^{-1}2^{\im(s;w)}+(\sm_1\sj_2)^{-1}2^{\im(s;w,u)}}~.~\frac{1}{1+\sj_2\sm_2^{-1}2^{-\im(y;u|w)}}\right.\n
                         &&~~~~~~\qquad\qquad\qquad\qquad\times{\mathbf{1}\left\{\mathsf{d}_1\left(s,\hat{s}_1(w)\right)\le \mathsf{D}_1,\mathsf{d}_2\left(s,\hat{s}_2(w,u,y)\right)\le \mathsf{D}_2\right\}}\Bigg)\label{eq:HB204}\\ 
                        &=&\e\dfrac{{\mathbf{1}\left\{\mathsf{d}_1\left(S,\hat{s}_1(W)\right)\le \mathsf{D}_1,\mathsf{d}_2\left(S,\hat{s}_2(W,U,Y)\right)\le \mathsf{D}_2\right\}}}{\left({1+\sm_1^{-1}2^{\im(S;W)}+(\sm_1\sj_2)^{-1}2^{\im(S;W,U)}}\right)
                        \left({1+\sj_2\sm_2^{-1}2^{-\im(Y;U|W)}}\right)}\label{eq:HB205}
                         \end{IEEEeqnarray}
                         
where \eqref{eq:HB201} is due to symmetry, the \underline{main step} \eqref{eq:HB202} follows from Jensen inequality for the 
 two-valued convex function $f(x_1,x_2)=\dfrac{1}{x_1x_2}$ on $\mathbb{R}^2_+$.  Equation \eqref{eq:HB203} follows from computing the denominator terms. This is discussed below. Lastly, equation \eqref{eq:HB204} follows from the facts that random codebook and random binning are independent and  $(W(1),U(1,1))$ is  generated according to $q_{WU}$.


\subsubsection*{Derivation of equation \eqref{eq:HB203}} Deriving this equation involves computing two denominator terms. To compute  $\e_{\mc,\mb|W(1),U(1,1),B(1)}\sum_{\bar{j}_2}2^{\im(y;U(1,\bar{j}_2)|W(1))}\mathbf{1}(B(\bar{j}_2)=1)$  for any $\bar{j}_2\neq 1$ we have:
\begin{align}
 &\e_{\mc,\mb|W(1),U(1,1),B(1)} 2^{\im(y;U(1,\bar{j}_2)|W(1))}\mathbf{1}(B(\bar{j}_2)=1)=\sum_uq(u|W(1))2^{\im(y;U(1,\bar{j}_2)|W(1))}\sm_2^{-1}\n&\qquad\qquad\qquad\qquad\qquad\qquad\qquad\qquad\qquad\qquad~~~=\sm_2^{-1}\sum_uq(u|W(1),y)=\sm_2^{-1},\label{eq:HB203-3}
\end{align}
where we have used similar argument used in the proof of \eqref{eq:BT203-1}. 

To compute $\e_{\mc|W(1),U(1,1)}\sum_{\tilde{m}_1,\tilde{j}_2}2^{\im(s;W(\tilde{m}_1),U(\tilde{m}_1,\tilde{j}_2))}$ we consider the following three cases separately:
\begin{itemize}
\item Case 1, $(\tilde{m}_1=1, \tilde{j}_2\neq 1)$:
\be
\e_{\mc|W(1),U(1,1)}2^{\im(s;W(1),U(1,\tilde{j}_2))}=\sum_uq(u|W(1))2^{\im(s;W(1),u)}=\sum_u\frac{q(W(1),u|s)}{q(W(1))}=2^{\im(s;W(1))},\label{eq:HB203-1}
\ee
where  we have used the fact that $U(1,\tilde{j}_2),\tilde{j}_2\neq 1$ is independent of  $W(1),U(1,1)$ and drawn from $q_{U|W}(.|W(1))$.
\item Case 2, $\tilde{m}_1\neq 1$:
\be\e_{\mc|W(1),U(1,1)}2^{\im(s;W(\tilde{m}_1),U(\tilde{m}_1,\tilde{j}_2))}=1,\label{eq:HB203-2}\ee
where we have used the fact that $U(\tilde{m}_1,\tilde{j}_2),\tilde{m}_1\neq 1$ is independent of  $W(1),U(1,1)$ and drawn from $q_{WU}$.
\item Case 3, $(\tilde{m}_1= 1, \tilde{j}_2= 1)$:
\be\e_{\mc|W(1),U(1,1)}2^{\im(s;W(1),U(1,1)}=2^{\im(s;W(1),U(1,1)}.\ee
\end{itemize}

\end{IEEEproof}
\subsection{Multiple Description}
 \begin{figure}\begin{center}
  \begin{tikzpicture}[scale=1.4,>=
   stealth'
  ]
\def\bscale{1.5}
\tikzstyle{enc}=[scale=.8,draw=black, text width=2.4em, rounded corners,
    text centered, minimum height=2.5em
    ]
    \tikzstyle{ch}=[scale=.8,draw=black, minimum width=2.6em, rounded corners,
    text centered, minimum height=6em,draw=blue,
    ]
\tikzstyle{ann} = [above, text width=2em]
\tikzstyle{dec} = [enc, text width=2.4em, 
    minimum height=2em,fill=gray!70!white,draw=black]
 \node (d0) [enc,scale=\bscale]{$\psi_0$};
 \path (d0.north)+(0,1) node (d1) [enc,scale=\bscale]{$\psi_1$};
  \path (d0.south)+(0,-1) node (d2) [enc,scale=\bscale]{$\psi_2$}; 
   \path (d0.west)+(-2,0) node (e0) [enc,scale=\bscale,minimum height=11.6em]{$\varphi$}; 
  \path (e0.west)+(-.5,0) node (e01) [scale=1]{$S$};
   \path (d0.east)+(1,0) node (d00) [scale=1]{$\widehat{S}_0,\mathsf{D}_0$};
    \path (d1.east)+(1,0) node (d10) [scale=1]{$\widehat{S}_1,\mathsf{D}_1$};
     \path (d2.east)+(1,0) node (d20) [scale=1]{$\widehat{S}_2,\mathsf{D}_2$};
  \path[draw,->] (e0.71)|- node(m)[above,near end]{$M_1$}(d1);
    \path[draw,->] (e0.-71)|- node(m2)[below,near end]{$M_2$}(d2);
   \draw[->](m)|-(d0.-200); 
         \draw[->](m2)|-(d0.200);
         \draw[->](e01)--(e0); 
           \draw[->](d0)--(d00);           \draw[->](d1)--(d10);
           \draw[->](d2)--(d20);
  \end{tikzpicture} 
\caption{Multiple description coding.}\label{fig:MD}
\end{center}

  \end{figure}
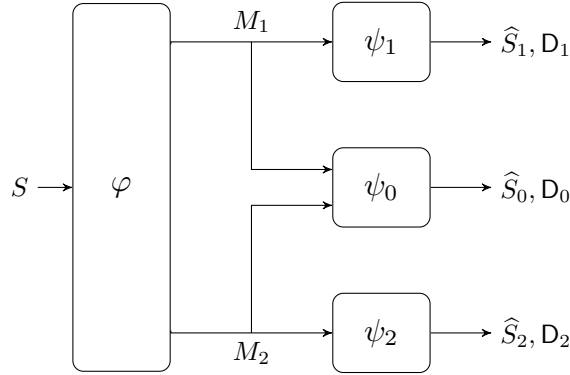
  
  Consider the problem of  multiple description coding  depicted in Fig. \ref{fig:MD}. Let $q_{S}$ be the  distribution of the source and $\mathsf{d}_k(s,\hat{s}_k), k=0,1,2$ be three distortion measures. We  will use the \emph{probability of excess distortion} as the criterion for measuring the reliability of the system.

 \begin{definition}
 An $(\sm_1,\sm_2)$-code for the multiple description coding  consists of (possibly stochastic) encoder $\varphi:\ms\mapsto[1:\sm_1]\times[1:\sm_2]$, and  (possibly stochastic) decoders $\psi_0:[1:\sm_1]\times[1:\sm_2]\mapsto\widehat{S}_0$, $\psi_1:[1:\sm_1]\mapsto\widehat{S}_1$ and $\psi_2:[1:\sm_2]\mapsto\widehat{S}_2$. Given a tuple of distortion levels $(\mathsf{D}_0,\mathsf{D}_1,\mathsf{D}_2)$, the probability of excess distortion is defined by,
\[\mathsf{P}[\me;\mathsf{D}_0,\mathsf{D}_1,\mathsf{D}_2]:=\mathsf{P}[\mathsf{d}_0(S,\widehat{S}_0)>\mathsf{D}_0\ \vee\ \mathsf{d}_1(S,\widehat{S})>\mathsf{D}_1\ \vee\ \mathsf{d}_2(S,\widehat{S}_2)>\mathsf{D}_2].\] 
Also, we define the probability of \emph{no-excess distortion} by 
\[\mathsf{P}[C;\mathsf{D}_0,\mathsf{D}_1,\mathsf{D}_2]:=\mathsf{P}[\mathsf{d}_0(S,\widehat{S}_0)\leq \mathsf{D}_0,\mathsf{d}_1(S,\widehat{S}_1)\leq \mathsf{D}_1,\  \mathsf{d}_2(S,\widehat{S}_2)\leq \mathsf{D}_2].\] 
 \end{definition}
 
 We prove a one-shot version of an equivalent characterization of Zhang-Berger inner bound \cite{ZB,cuffmd} for this problem. 
 
  \begin{theorem}
 Given any  $q_{U_0U_1U_2|S}$ and functions $\hat{s}_0(u_0,u_1,u_2)$, $\hat{s}_1(u_0,u_1)$ and $\hat{s}_2(u_0,u_2)$, there is an  $(\sm_1,\sm_2)$-code for a single use of the source whose probability of \emph{no-excess distortion} is bounded from below by
 \be
\mathsf{P}[C;\mathsf{D}_0,\mathsf{D}_1,\mathsf{D}_2]\ge \e\frac{\mathbf{1}\left\{\mathsf{d}_0\left(S,\hat{s}_0(U_0,U_1,U_2)\right)\le \mathsf{D}_0, 
 \mathsf{d}_1\left(S,\hat{s}_1(U_0,U_1)\right)\le \mathsf{D}_1,
\mathsf{d}_2\left(S,\hat{s}_2(U_0,U_2)\right)\le \mathsf{D}_2\right\}}{1+\sj_0^{-1}2^{\im(S;U_0)}+\sm_1^{-1}2^{\im(S;U_0U_1)}+\sm_2^{-1}2^{\im(S;U_0U_2)}+\sj_0(\sm_1\sm_2)^{-1}2^{\im(S;U_0U_1U_2)+\im(U_1;U_2|U_0)}},
 \ee
  where $\sj_0$ is a common divisor of $\sm_1$ and $\sm_2$.
  Moreover, loosening this bound gives the following upper bound on the probability of excess distortion 
of the code,
 \begin{align}
 \mathsf{P}&\left[\mathsf{d}_0\left(S,\hat{s}_0(U_0,U_1,U_2)\right)> \mathsf{D}_0, \mathsf{or}~
 \mathsf{d}_1\left(S,\hat{s}_1(U_0,U_1)\right)> \mathsf{D}_1, \mathsf{or}
 ~\mathsf{d}_2\left(S,\hat{s}_2(U_0,U_2)\right)> \mathsf{D}_2
 , \mathsf{or}\ \right. \n
 &~~~\log\sj_0-\im(S;U_0)<\gamma, \mathsf{or}\ ~ \log\sm_1-\im(S;U_0U_1)<\gamma, \mathsf{or}\ ~ \log\sm_2-\im(S;U_0U_2)<\gamma, \mathsf{or}\n& ~~~~ \log\frac{\sm_1\sm_2}{\sj_0}-\im(S;U_0U_1U_2)-\im(U_1;U_2|U_0)<\gamma\Big]+4\times2^{-\gamma}.
\end{align}
  \end{theorem}

\begin{IEEEproof} 
 We only prove the lower bound on probability of no-excess distortion. Derivation of the loosened bound is similar to that of Gelfand-Pinsker and thus omitted. Since $\sj_0$ is a common divisor of $\sm_1$ and $\sm_2$, there exists positive integers $\sj_1$ and $\sj_2$ such that $\sm_1=\sj_0\sj_1$ and $\sm_2=\sj_0\sj_2$. 
 \subsubsection*{Random codebook generation}
 \begin{itemize}
 \item $\mc_0=\{U_0(j_0)\}_{j_0=1}^{\sj_0}$ is a random codebook whose elements are drawn independently from $q_{U_0}$.
 \item For each $j_0\in[1:\sj_0]$, let $\mc_{12}(j_0)=\mc_1(j_0)\times\mc_2(j_0)=\{U_1(j_1)\}_{j_1=1}^{\sj_1}\times\{U_2(j_2)\}_{j_2=1}^{\sj_2}$ be a random product codebook, in which the codebooks $\mc_1(j_0)$ and $\mc_2(j_0)$ are generated independently and the elements of the codebook $\mc_k(j_0),k=1,2$   are drawn independently from $q_{U_k|U_0}(.|U_0(j_0))$. Moreover the codebooks $\mc_{12}(j_0),j_0\in[1:\sj_0]$ are generated independently.
 \item Let $\mc$ be the set of all codewords. 
 \end{itemize}
 
 \subsubsection*{Encoding} Encoder  uses an  SMC for three r.v.'s  to obtain the tuple $(j_0,j_1,j_2)$. Given $s$, the SMC chooses a tuple $(j_0,j_1,j_2)$ with the probability
\[
P_{\enc}(j_0,j_1,j_2|s)=\dfrac{2^{\im_q(s;U_0(j_0),U_1(j_0,j_1),U_2(j_0,j_2))+\im_q(U_1(j_0,j_1);U_2(j_0,j_2)|U_0(1))}}{\sum_{\tilde{j}_0,\tilde{j}_1,\tilde{j}_2}2^{\im_q(s;U_0(\tilde{j}_0),U_1(\tilde{j}_0,\tilde{j}_1),U_2(\tilde{j}_0,\tilde{j}_2))+\im_q(U_1(\tilde{j}_0,\tilde{j}_1);U_2(\tilde{j}_0,\tilde{j}_2)|U_0(\tilde{j}_0))}}.
\]  
Then the encoder sends $m_1=(j_0,j_1)$ and $m_2=(j_0,j_2)$ to the decoders. 
\subsubsection*{Decoding} Observing $(m_1,m_2)=(j_0,j_1,j_2)$, decoder 0 computes $\hat{s}_0(U_0(j_0),U_1(j_0,j_1),U_2(j_0,j_2))$ as the estimate of $S$. Observing $m_k=(j_0,j_k)$, decoder $k=1,2$ computes $\hat{s}_0(U_0(j_0),U_k(j_0,j_k))$ as the estimate of $S$.
\subsubsection*{Analysis} Observe that the joint distribution of random variables factors as,
\begin{align*}
P(s,j_0,j_1,j_{2})=q(s)P_{\enc}(j_0,j_1,j_2|s).
\end{align*}
  The probability of no-excess distortion can be bounded from below by 
\begin{align*}\mathsf{P}[C]\ge\sum_{s,y,m_{1:2},j_{2}}&P(s,j_0,j_1,j_{2})\chi_{s,U_0(j_0),U_1(j_0,j_1),U_2(j_0,j_2)},
\end{align*} 
where 
\[\chi_{s,u_0,u_1,u_2}:=\mathbf{1}\left\{\mathsf{d}_0\left(S,\hat{s}_0(u_0,u_1,u_2)\right)> \mathsf{D}_0, 
 \mathsf{d}_1\left(S,\hat{s}_1(u_0,u_1)\right)> \mathsf{D}_1,
\mathsf{d}_2\left(S,\hat{s}_2(u_0,u_2)\right)> \mathsf{D}_2\right\}.\]
Skipping similar symmetry arguments we have:

 \begin{IEEEeqnarray}{rCl}
&\e&\mathsf{P}[C]\ge
                         \e_{\mc}\sum_{s}\sj_0\sj_1\sj_2q(s)\dfrac{2^{\im(s;U_0(1),U_1(1,1),U_2(1,1))+\im(U_1(1,1);U_2(1,1)|U_0(1))}}{\sum_{\tilde{j}_0,\tilde{j}_1,\tilde{j}_2}2^{\im(s;U_0(\tilde{j}_0),U_1(\tilde{j}_0,\tilde{j}_1),U_2(\tilde{j}_0,\tilde{j}_2))+\im(U_1(\tilde{j}_0,\tilde{j}_1);U_2(\tilde{j}_0,\tilde{j}_2)|U_0(\tilde{j}_0))}}\chi_{s,U_0(1),U_1(1,1),U_2(1,1)}\n
                         &\ge&\sum_{s}q(s)\textcolor{blue!90!black}{\e_{U_{0:2}(1,1,1)}}\dfrac{\textcolor{blue!90!black}{\sj_0\sj_1\sj_2 2^{\im(s;U_0(1),U_1(1,1),U_2(1,1))+\im(U_1(1,1);U_2(1,1)|U_0(1))}\chi_{s,U_0(1),U_1(1,1),U_2(1,1)}}}{\textcolor{red!60!blue}{\e_{\mc|U_0(1),U_1(1,1),U_2(1,1)}\sum_{\tilde{j}_0,\tilde{j}_1,\tilde{j}_2}2^{\im(s;U_0(\tilde{j}_0),U_1(\tilde{j}_0,\tilde{j}_1),U_2(\tilde{j}_0,\tilde{j}_2))+\im(U_1(\tilde{j}_0,\tilde{j}_1);U_2(\tilde{j}_0,\tilde{j}_2)|U_0(\tilde{j}_0))}}}\label{eq:ZB202}\\
                        &\ge&\sum_{s}q(s){\e_{U_{0:2}(1,1,1)}}{{\sj_0\sj_1\sj_2 2^{\im(s;U_0(1),U_1(1,1),U_2(1,1))+\im(U_1(1,1);U_2(1,1)|U_0(1))}\chi_{s,U_0(1),U_1(1,1),U_2(1,1)}}}\n
                        &&~~~~~~\times\left(\sj_0\sj_1\sj_2+\sj_1\sj_22^{\im(s;U_0(1))}+\sj_22^{\im(s;U_0(1),U_1(1,1))}+\sj_12^{\im(s;U_0(1),U_2(1,1))}\right.\n
                        &&~~~~~~~~~~~~~\qquad\qquad\qquad~~~~~\left.+2^{\im(s;U_0(1),U_1(1,1),U_2(1,1))+\im(U_1(1,1);U_2(1,1)|U_0(1))}\right)^{-1}\label{eq:ZB203}\\                      
                         &=&\sum_{s,u_0,u_1,u_2}q(s,u_0,u_1,u_2)\frac{\chi_{s,u_0,u_1,u_2}}{1+\sj_0^{-1}2^{\im(s;u_0)}+(\sj_0\sj_1)^{-1}2^{\im(s;u_0u_1)}+(\sj_0\sj_2)^{-1}2^{\im(s;u_0u_2)}+(\sj_0\sj_1\sj_2)^{-1}2^{\im(s;u_0u_1u_2)+\im(u_1;u_2|u_0)}}\label{eq:ZB204}\\ 
                        &=&\e\frac{\mathbf{1}\left\{\mathsf{d}_0\left(S,\hat{s}_0(U_0,U_1,U_2)\right)> \mathsf{D}_0, 
 \mathsf{d}_1\left(S,\hat{s}_1(U_0,U_1)\right)> \mathsf{D}_1,
\mathsf{d}_2\left(S,\hat{s}_2(U_0,U_2)\right)> \mathsf{D}_2\right\}}{1+\sj_0^{-1}2^{\im(S;U_0)}+\sm_1^{-1}2^{\im(S;U_0U_1)}+\sm_2^{-1}2^{\im(S;U_0U_2)}+\sj_0(\sm_1\sm_2)^{-1}2^{\im(S;U_0U_1U_2)+\im(U_1;U_2|U_0)}}\label{eq:ZB205}
                         \end{IEEEeqnarray}

where the \underline{main step} \eqref{eq:ZB202} follows from Jensen inequality for the 
 two-valued convex function $f(x)=\dfrac{1}{x}$ on $\mathbb{R}_+$.  
Equation \eqref{eq:ZB203} follows from separating the cases ..... This is discussed below. Lastly, equation \eqref{eq:ZB204}  follows from the facts that $(U_0(1),U_1(1,1),U_2(1,1))$ is  generated according to $q_{U_0}q_{U_1|U_0}q_{U_2|U_0}$.

\subsubsection*{Derivation of equation \eqref{eq:ZB203}}
 In the following we consider each of the three case mentioned above separately. But first observe that 
 \[2^{\im(s;u_0,u_1,u_2)+\im(u_1;u_2|u_0)}=\dfrac{q(u_0,u_1,u_2)}{q(u_0)q(u_1|u_0)q(u_2|u_0)}.\]

\begin{itemize}
\item Case 1, $\tilde{j_0}\neq 1$:
\begin{align}
\e_{\mc|U_0(1),U_1(1,1),U_2(1,1)}&2^{\im(s;U_0(\tilde{j}_0),U_1(\tilde{j}_0,\tilde{j}_1),U_2(\tilde{j}_0,\tilde{j}_2))+\im(U_1(\tilde{j}_0,\tilde{j}_1);U_2(\tilde{j}_0,\tilde{j}_2)|U_0(\tilde{j}_0))}\n&=\sum_{u_0,u_1,u_2}q(u_0)q(u_1|u_0)q(u_2|u_0)2^{\im(s;u_0,u_1,u_2)+\im(u_1;u_2|u_0)}\n&=1,\label{eq:ZB203-1}
\end{align}
where  we have used the fact that $(U_0(\tilde{j}_0),U_1(\tilde{j}_0,\tilde{j}_1),U_2(\tilde{j}_0,\tilde{j}_2)),\tilde{j}_0\neq 1$ is independent of  $(U_0(1),U_1(1,1),U_2(1,1))$ and drawn from $q_{U_0}q_{U_1|U_0}q_{U_2|U_0}$.
\item Case 2, $\tilde{j}_0=1$ and $\tilde{j_1}\neq 1,\tilde{j_2}\neq 1$:
\begin{align}
\e_{\mc|U_0(1),U_1(1,1),U_2(1,1)}&2^{\im(s;U_0(1),U_1(1,\tilde{j}_1),U_2(1,\tilde{j}_2))+\im(U_1(1,\tilde{j}_1);U_2(1,\tilde{j}_2)|U_0(1))}\n&=\sum_{u_1,u_2}q(u_1|U_0(1))q(u_2|U_0(1))2^{\im(s;U_0(1),u_1,u_2)+\im(u_1;u_2|U_0(1))}\n&=\sum_{u_1,u_2}\frac{q(U_0(1),u_1,u_2|s)}{q(U_0(1))}\n&=2^{\im(s;U_0(1))},\label{eq:ZB203-2}
\end{align}
where we have used the fact that given $U_0(1)$, $(U_1(1,\tilde{j}_1),U_2(1,\tilde{j}_2)),\tilde{j_1}\neq 1,\tilde{j_2}\neq 1$ is independent of  $(U_1(1,1),U_2(1,1))$ and drawn from $q_{U_1|U_0}(.|U_0(1))q_{U_2|U_0}(.|U_0(1))$. 
\item Case 3, $\tilde{j}_0=\tilde{j}_1=1$ and $\tilde{j_2}\neq 1$: Similarly we have,
\begin{align}
\e_{\mc|U_0(1),U_1(1,1),U_2(1,1)}&2^{\im(s;U_0(1),U_1(1,1),U_2(1,\tilde{j}_2))+\im(U_1(1,1);U_2(1,\tilde{j}_2)|U_0(1))}\n
&=2^{\im(s;U_0(1),U_1(1,1))}\label{eq:ZB203-3}
\end{align}
\item Case 4, $\tilde{j}_0=\tilde{j}_2=1$ and $\tilde{j_1}\neq 1$: Similarly we have,
\begin{align}
\e_{\mc|U_0(1),U_1(1,1),U_2(1,1)}&2^{\im(s;U_0(1),U_1(1,\tilde{j}_1),U_2(1,1))+\im(U_1(1,\tilde{j}_1);U_2(1,1)|U_0(1))}\n&=2^{\im(s;U_0(1),U_2(1,1))}\label{eq:ZB203-4}
\end{align}
\item Case 5, $(\tilde{j}_0,\tilde{j}_1,\tilde{j}_2)=(1,1,1)$:
\begin{align}
\e_{\mc|U_0(1),U_1(1,1),U_2(1,1)}&2^{\im(s;U_0(1),U_1(1,1),U_2(1,1))+\im(U_1(1,1);U_2(1,1)|U_0(1))}\n&=2^{\im(s;U_0(1),U_1(1,1),U_2(1,1))+\im(U_1(1,1);U_2(1,1)|U_0(1))}.
\end{align}
\end{itemize}

Applying \eqref{eq:ZB203-1}-\eqref{eq:ZB203-4} to \eqref{eq:ZB202} yields \eqref{eq:ZB203}. 
\end{IEEEproof} 
\section{Joint source-channel coding}
\label{JSCC}
To illustrate the application of our technique to joint source-channel coding problems, we study the problem of joint source-channel coding over MAC in this section.
\subsection{Joint source-channel coding over MAC}
  \begin{figure}\begin{center}\begin{tikzpicture}[scale=1.4,>=
   stealth'
  ]
\def\bscale{1.5}
\tikzstyle{enc}=[scale=.8,draw=black, minimum width=2.9em, rounded corners,
    text centered, minimum height=2.1em
    ]
    \tikzstyle{ch}=[scale=.8,draw=black, minimum width=2.6em, rounded corners,
    text centered, minimum height=6em,draw=blue,
    ]
\tikzstyle{ann} = [above, text width=2em]
\tikzstyle{dec} = [enc, text width=2.4em, 
    minimum height=2em,fill=gray!70!white,draw=black]
 \node (d0) [coordinate]{};
 \path (d0.north)+(0,1) node (d1) [enc,scale=\bscale]{$\varphi_1$};
  \path (d0.south)+(0,-1) node (d2) [enc,scale=\bscale]{$\varphi_2$}; 
   \path (d0.east)+(2,0) node (e0) [enc,scale=\bscale,minimum height=9em]{$q_{Y|X_1X_2}$}; 
     \path (e0.0)+(1,0) node (dec) [enc,scale=\bscale]{$\psi$}; 
    \path (d1.west)+(-1,0) node (d10) [scale=1.2]{${S}_1$};
     \path (d2.west)+(-1,0) node (d20) [scale=1.2]{${S}_2$};
          \path (dec)+(1.5,0) node (es) [scale=1.2]{$\widehat{S}_1,\widehat{S}_2$};
  \path[draw,<-] (e0.122)-- node(m)[above]{$X_1$}(d1);
    \path[draw,<-] (e0.-122)-- node(m2)[below]{$X_2$}(d2);
      \path[draw,->] (e0)-- node[above]{$Y$}(dec);
          \draw[<-](d1)--(d10);
           \draw[<-](d2)--(d20);
           \draw[->](dec)--(es);
  \end{tikzpicture} 
 \caption{Joint source-channel coding over MAC.}\label{fig:smac}
\end{center}

 \end{figure}
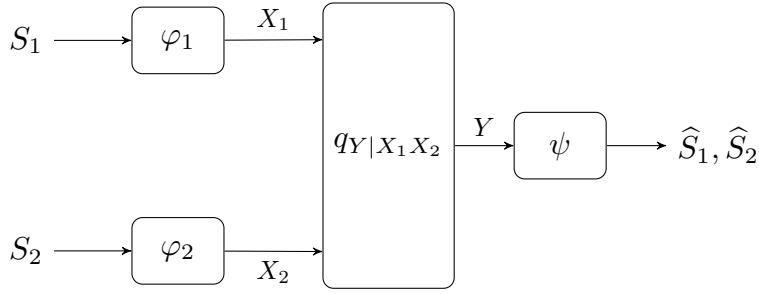 
Consider the problem of transmission of correlated sources $(S_1,S_2)$ over a MAC depicted in Fig. \ref{fig:smac}. Let $q_{Y|X_1X_2}$ be channel transition probability.
 We prove a one-shot version of the result of Cover-El Gamal-Salehi \cite{ceg} for this problem.
  \begin{definition}
 An code for lossless transmission of correlated source over a MAC consists of (possibly stochastic) encoders $\varphi_k:\ms_k\mapsto\mx_k, k=1,2$, and  (possibly stochastic) decoder $\psi:\my\mapsto\ms_1\times\ms_2$. 
 \end{definition} 
\begin{theorem}
Let $K$ be the common part of $S_1$ and $S_2$. Given any $q(s_1,s_2,t,x_1,x_2)=q(s_1,s_2)q(t)q(x_1|s_1,t)q(x_2|s_2,t)$, there is a code for a single use of the channel whose probability of correct decoding is bounded from below by
\begin{IEEEeqnarray*}{lCl}
\e \left(1+2^{h(S_1|S_2)-\im(Y;X_1|X_2,S_2,T)}+
                         2^{h(S_2|S_1)-\im(Y;X_2|X_1,S_1,T)}+2^{h(S_1,S_2|K)-\im(Y;X_1,X_2|K,T)}+2^{h(S_1,S_2)-\im(Y;X_1,X_2)}\right)^{-1}.
\end{IEEEeqnarray*}

Moreover, loosening this bound gives the following upper bound on error probability of the code,
\begin{align}\p[&\im(Y;X_1|X_2,S_2,T)-h(S_1|S_2)<\gamma\ ,\quad ~~\mathsf{or}\n& \im(Y;X_2|X_1,S_1,T)-h(S_2|S_1)<\gamma\ ,\quad ~~\mathsf{or}\n& \im(Y;X_1,X_2|K,T)-h(S_1,S_2|K)<\gamma\ ,\ ~\mathsf{or}\n& \im(Y;X_1,X_2)-h(S_1,S_2)<\gamma]+4\times2^{-\gamma},\label{eq:300}\end{align}
where $\gamma$ is any positive number.
\end{theorem} 
\begin{remark} Second order analysis of the above bound can be obtained in a similar manner as the Gelfand-Pinsker rate region. We will include it in the next version of this draft. Its equations resemble the four constraints given by \cite{ceg}. It subsumes the result of \cite{Tan} whose region includes two additional constraints.    
\end{remark}

\begin{IEEEproof}
We only prove the lower bound on probability of correct decoding. Derivation of the loosened bound is similar to that of Gelfand-Pinsker and is thus omitted.
\subsubsection*{Codebook generation}
We employ a one-shot version of the codebook used in \cite{ceg}. 
\begin{itemize}
\item For each $k\in\mk$, draw a symbol $T(k)$ from pmf $q(t)$,
\item For each $s_1$, draw a symbol $X_1(s_1)$ from $q\big(x_1|s_1,T(k(s_1))\big)$,
\item For each $s_2$, draw a symbol $X_2(s_2)$ from $q\big(x_2|s_2,T(k(s_2))\big)$.
\end{itemize}

\subsubsection*{Encoding} Given $s_j$, transmitter $j$ sends $X_j(s_j)$. 
\subsubsection*{Decoding} In contrast to Gelfand-Pinsker and broadcast channel, we use an SLC for decoding.  Observing $y$, decoder uses the SLC $P_{S_1S_2|Y}(\hat{s}_1,\hat{s}_2|y)$ to find  an estimate of $(s_1,s_2)$. The SLC can be formulated as follows:
\small
 \begin{align*}
P_{S_1S_2|Y}(\hat{s}_1,\hat{s}_2|y)&=\dfrac{q(\hat{s}_1,\hat{s}_2)q(y|X_1(\hat{s}_1),X_2(\hat{s}_2))}{\sum_{\bar{s}_1,\bar{s}_2}q(\bar{s}_1,\bar{s}_2)q(y|X_1(\bar{s}_1),X_2(\bar{s}_2))}\\
                                                        &=\dfrac{q(\hat{s}_1,\hat{s}_2)2^{\im_q(y;X_1(\hat{s}_1),X_2(\hat{s}_2))}}{\sum_{\bar{s}_1,\bar{s}_2}q(\bar{s}_1,\bar{s}_2)2^{\im_q(y;X_1(\bar{s}_1),X_2(\bar{s}_2))}}.
\end{align*} 
\normalsize
\subsubsection*{Analysis} Observe that the joint distribution of random variables factors as,
\begin{align*}
P(s_1,s_2,y,\hat{s}_{1},\hat{s}_{2})=q(s_1,s_2)q\big(y|X_1(s_1),X_2(s_2)\big)P_{S_1S_2|Y}(\hat{s}_1,\hat{s}_2|y).
\end{align*}
The probability of correct decoding can be bounded from below by $\mathsf{P}[C]\ge\sum_{s_1,s_2,y}P(s_1,s_2,y,\hat{s}_{1}=s_1,\hat{s}_{2}=s_2),$  hence we have:

\begin{IEEEeqnarray}{rCl}
\e\mathsf{P}[C]&=&\e\sum_{s_1,s_2,y}q(s_1,s_2)q(y|(X_1(s_1),X_2(s_2))\dfrac{q(s_1,s_2)2^{\im(y;X_1(s_1),X_2(s_2))}}{\sum_{\bar{s}_1,\bar{s}_2}q(\bar{s}_1,\bar{s}_2)2^{\im(y;X_1(\bar{s}_1),X_2(\bar{s}_2))}}\\
&=&\e\sum_{s_1,s_2,y:\atop (s_1,s_2)\in\mathsf{Supp}}q(s_1,s_2)q(y|(X_1(s_1),X_2(s_2))\dfrac{q(s_1,s_2)2^{\im(y;X_1(s_1),X_2(s_2))}}{\sum_{\bar{s}_1,\bar{s}_2}q(\bar{s}_1,\bar{s}_2)2^{\im(y;X_1(\bar{s}_1),X_2(\bar{s}_2))}}\\
                         &\ge&\sum_{s_1,s_2,y:\atop (s_1,s_2)\in\mathsf{Supp}}q(s_1,s_2)\textcolor{blue!90!black}{\e_{T(k),X_1(s_1),X_2(s_2)}q(y|(X_1(s_1),X_2(s_2))}
                      \dfrac{\textcolor{blue!90!black}{q(s_1,s_2)2^{\im(y;X_1(s_1),X_2(s_2))}}}{\textcolor{red!60!blue}{\e_{\mc|T(k),X_1(s_1),X_2(s_2)}\sum_{\bar{s}_1,\bar{s}_2}q(\bar{s}_1,\bar{s}_2)2^{\im(y;X_1(\bar{s}_1),X_2(\bar{s}_2))}}}\label{eq:js102}
\IEEEeqnarraynumspace\end{IEEEeqnarray}
where  $\mathsf{Supp}$ is the support set of the pmf $q_{S_1S_2}$, that is, $\mathsf{Supp}=\{(s_1,s_2):q_{S_1S_2}(s_1,s_2)\neq 0\}$. By 
 $T(k)$ we mean $T(k(s_1))=T(k(s_2))$, and the \underline{main step} \eqref{eq:js102} follows from the Jensen inequality for the  convex function $f(x)=\dfrac{1}{x}$ on $\mathbb{R}_+$. To evaluate \eqref{eq:js102}, we first find an upper bound on the denominator. To do this, we first split the denominator into five terms as follows,
\begin{IEEEeqnarray}{rCl}
\e_{\mc|T(k),X_1(s_1),X_2(s_2)}\sum_{\bar{s}_1,\bar{s}_2}q(\bar{s}_1,\bar{s}_2)2^{\im(y;X_1(\bar{s}_1),X_2(\bar{s}_2))}&=&\sum_{(\bar{s}_1,\bar{s}_2)\in\mathsf{Supp}:\atop {k}(\bar{s}_1)=k(\bar{s}_2)\neq k}q(\bar{s}_1,\bar{s}_2)\e_{\mc|T(k),X_1(s_1),X_2(s_2)}2^{\im(y;X_1(\bar{s}_1),X_2(\bar{s}_2))}\n
&+&\sum_{(\bar{s}_1,\bar{s}_2)\in\mathsf{Supp}:\atop \bar{s}_1\neq s_1,\bar{s}_2\neq s_2, {k}(\bar{s}_1)={k}(\bar{s}_2)= k}q(\bar{s}_1,\bar{s}_2)\e_{\mc|T(k),X_1(s_1),X_2(s_2)}2^{\im(y;X_1(\bar{s}_1),X_2(\bar{s}_2))}\n
&+&\sum_{\bar{s}_1:(\bar{s}_1,{s}_2)\in\mathsf{Supp}:\atop \bar{s}_1\neq s_1}q(\bar{s}_1,{s}_2)\e_{\mc|T(k),X_1(s_1),X_2(s_2)}2^{\im(y;X_1(\bar{s}_1),X_2({s}_2))}\n
&+&\sum_{\bar{s}_2:({s}_1,\bar{s}_2)\in\mathsf{Supp}:\atop \bar{s}_2\neq s_2}q({s}_1,\bar{s}_2)\e_{\mc|T(k),X_1(s_1),X_2(s_2)}2^{\im(y;X_1({s}_1),X_2(\bar{s}_2))}\n
&+&q({s}_1,{s}_2)2^{\im(y;X_1({s}_1),X_2({s}_2))}.\label{eq:js0}
\end{IEEEeqnarray}
We find upper bounds on each term of \eqref{eq:js0} separately.
\begin{itemize}
\item Case 1:
$\bar{k}\neq k(=k(s_1)=k(s_2))$, where $\bar{k}=k(\bar{s}_1)=k(\bar{s}_2)$,
\begin{IEEEeqnarray}{rCl}
\e_{\mc|T(k),X_1(s_1),X_2(s_2)}2^{\im(y;X_1(\bar{s}_1),X_2(\bar{s}_2))}&=&\sum_{t,x_1,x_2}q(t)q(x_1|\bar{s}_1,t)q(x_2|\bar{s}_2,t)2^{\im(y;x_1,x_2)}\label{eq:js300}\\
                                                                                                         &=&\sum_{t,x_1,x_2}q(t,x_1,x_2|\bar{s}_1,\bar{s}_2)\dfrac{q(y|x_1,x_2)}{q(y)}\label{eq:js301}\\
                                                                                                         &=&\dfrac{\sum_{t,x_1,x_2}q(t,x_1,x_2,y|\bar{s}_1,\bar{s}_2)}{q(y)}\label{eq:js302}\\
                                                                                                         &=&\dfrac{q(y|\bar{s}_1,\bar{s}_2)}{q(y)},\label{eq:js303}
\end{IEEEeqnarray}
where \eqref{eq:js300} and \eqref{eq:js301} follows from the fact that $(T(\bar{k}),X_1(\bar{s}_1),X_2(\bar{s}_2))$ is independent of $(T({k}),X_1({s}_1),X_2({s}_2))$ and is drawn from $q_Tq_{X_1|S_1T}(.|\bar{s}_1,.)q_{X_2|S_2T}(.|\bar{s}_2,.)=q_{TX_1X_2|S_1,S_2}(.|\bar{s}_1,\bar{s}_2)$. Using \eqref{eq:js303} we obtain the following upper bound on the first term of \eqref{eq:js0}: 
\begin{IEEEeqnarray}{rCl}
\sum_{(\bar{s}_1,\bar{s}_2)\in\mathsf{Supp}:\atop {k}\neq k}q(\bar{s}_1,\bar{s}_2)\e_{\mc|T(k),X_1(s_1),X_2(s_2)}2^{\im(y;X_1(\bar{s}_1),X_2(\bar{s}_2))}&\le&\sum_{(\bar{s}_1,\bar{s}_2)\in\mathsf{Supp}}\dfrac{q(\bar{s}_1,\bar{s}_2,y)}{q(y)}\n&=&1.
\end{IEEEeqnarray}

\item Case 2: $\bar{s}_1\neq s_1$, $\bar{s}_2\neq s_2$, but $\bar{k}=k(s_1)=k(s_2)=k$,
\begin{IEEEeqnarray}{rCl}
\e_{\mc|T(k),X_1(s_1),X_2(s_2)}2^{\im(y;X_1(\bar{s}_1),X_2(\bar{s}_2))}&=&\sum_{x_1,x_2}q(x_1|\bar{s}_1,T(k))q(x_2|\bar{s}_2,T(k))2^{\im(y;x_1,x_2)}\label{eq:js400}\\
                                                                                                         &=&\sum_{x_1,x_2}{q(x_1,x_2|\bar{s}_1,\bar{s}_2,T(k))}\dfrac{q(y|x_1,x_2)}{q(y)}\label{eq:js401}\\
                                                                                                         &=&\dfrac{\sum_{x_1,x_2}q(x_1,x_2,y|\bar{s}_1,\bar{s}_2,T(k))}{q(y)}\label{eq:js402}\\
                                                                                                         &=&\dfrac{q(y|\bar{s}_1,\bar{s}_2,T(k))}{q(y)},\label{eq:js403}
\end{IEEEeqnarray}
where \eqref{eq:js400} and \eqref{eq:js401} follows from the fact that given $T(k)$, $(X_1(\bar{s}_1),X_2(\bar{s}_2))$ is independent of $(X_1({s}_1),X_2({s}_2))$ and is drawn from $q_{X_1|S_1T}(.|\bar{s}_1,T(k))q_{X_2|S_2T}(.|\bar{s}_2,T(k))=q_{X_1X_2|S_1,S_2,T}(.|\bar{s}_1,\bar{s}_2,T(k))$. Using \eqref{eq:js403} we obtain the following upper bound on the second term of \eqref{eq:js0}:
\begin{IEEEeqnarray}{rCl}
\sum_{(\tilde{s}_1,\bar{s}_2)\in\mathsf{Supp}:\atop \bar{s}_1\neq s_1,\bar{s}_2\neq s_2,{k}(\bar{s}_1)= k}q(\bar{s}_1,\bar{s}_2)\e_{\mc|T(k),X_1(s_1),X_2(s_2)}2^{\im(y;X_1(\bar{s}_1),X_2(\bar{s}_2))}&\le&\sum_{(\tilde{s}_1,\bar{s}_2)\in\mathsf{Supp}:\atop {k}(\bar{s}_1)= k}q(\bar{s}_1,\bar{s}_2|T(k))\dfrac{q(y|\bar{s}_1,\bar{s}_2,T(k))}{q(y)}\IEEEeqnarraynumspace\label{eq:js404}\\
&=&\sum_{(\tilde{s}_1,\bar{s}_2)\in\mathsf{Supp}:\atop {k}(\bar{s}_1)= k}\dfrac{q(\bar{s}_1,\bar{s}_2,y|T(k))}{q(y)}\n
&=&\dfrac{q(k,y|T(k))}{q(y)}\label{eq:js406}\\
&=&q(k)\dfrac{q(y|k,T(k))}{q(y)}=q(k)2^{\im(y;k,T(k))}\label{eq:js407} 
\end{IEEEeqnarray}
where \eqref{eq:js404} is due to the fact that $T$ is independent of $(S_1,S_2)$ and \eqref{eq:js406} follows from the definition of common part, and $q(k)=\sum_{(\tilde{s}_1,\bar{s}_2)\in\mathsf{Supp}:\atop {k}(\bar{s}_1)= k}q(\bar{s}_1, \bar{s}_2)$.
\item Case 3: $\bar{s}_1\neq s_1$, $\bar{s}_2=s_2$, hence $\bar{k}=k(s_2)=k$,
\begin{IEEEeqnarray}{rCl}
\e_{\mc|T(k),X_1(s_1),X_2(s_2)}2^{\im(y;X_1(\bar{s}_1),X_2({s}_2))}&=&\sum_{x_1}q(x_1|\bar{s}_1,T(k))2^{\im(y;x_1,X_2({s}_2))}\label{eq:js500}\\
                                                                                                         &=&\sum_{x_1}{q(x_1|\bar{s}_1,{s}_2,T(k),X_2(s_2))}\dfrac{q(y|x_1,X_2({s}_2))}{q(y)}\label{eq:js501}\\
                                                                                                         &=&\dfrac{\sum_{x_1}q(x_1,y|\bar{s}_1,{s}_2,T(k),X_2(s_2))}{q(y)}\label{eq:js502}\\
                                                                                                         &=&\dfrac{q(y|\bar{s}_1,{s}_2,T(k),X_2(s_2))}{q(y)},\label{eq:js503}
\end{IEEEeqnarray}
where \eqref{eq:js500} and \eqref{eq:js501} follows from the fact that given $(T(k),X_2(s_2))$, $X_1(\bar{s}_1)$ is independent of $X_1({s}_1)$ and is drawn from $q_{X_1|S_1T}(.|\bar{s}_1,T(k))=q_{X_1X_2|S_1,S_2,T}(.|\bar{s}_1,s_2,T(k),X_2(s_2))$. Using \eqref{eq:js503} we obtain the following upper bound on the third term of \eqref{eq:js0}:

\begin{IEEEeqnarray}{rCl}
\sum_{\bar{s}_1:(\bar{s}_1,{s}_2)\in\mathsf{Supp} \atop \bar{s}_1\neq s_1}q(\bar{s}_1,{s}_2)&&\e_{\mc|T(k),X_1(s_1),X_2(s_2)}2^{\im(y;X_1(\bar{s}_1),X_2({s}_2))}\n&\le&\sum_{\bar{s}_1:(\bar{s}_1,{s}_2)\in\mathsf{Supp}}q({s}_2)q(\bar{s}_1|{s}_2,T(k))\dfrac{q(y|\bar{s}_1,{s}_2,T(k),X_2(s_2))}{q(y)}\label{eq:js504}\\
&\le&\sum_{\bar{s}_1:(\bar{s}_1,{s}_2)\in\mathsf{Supp}}q({s}_2)q(\bar{s}_1|{s}_2,T(k),X_2(s_2))\dfrac{q(y|\bar{s}_1,{s}_2,T(k),X_2(s_2))}{q(y)}\label{eq:js505}\\
&\le&\sum_{\bar{s}_1}q({s}_2)\dfrac{q(\bar{s}_1,y|{s}_2,T(k),X_2(s_2))}{q(y)}\\
&=&q({s}_2)\dfrac{q(y|{s}_2,T(k),X_2(s_2))}{q(y)}=q(s_2)2^{\im(y;s_2,T(k),X_2(s_2))},\label{eq:js506}
\end{IEEEeqnarray}
where \eqref{eq:js504} is due to the fact that $T$ is independent of $(S_1,S_2)$ in the pmf $q$ that we started with (this should be confused with the pmf induced by the code), and \eqref{eq:js505} follows from the Markov chain $S_1-S_2T-X_2$.
\item Case 4: $\bar{s}_2\neq s_2$, $\bar{s}_1=s_1$, hence ${k}=k(s_1)=k$,
Using symmetry between Case 3 and Case 4, we have the following bound on the fourth term of \eqref{eq:js0}:
\begin{IEEEeqnarray}{rCl}
\sum_{\bar{s}_1:({s}_1,\bar{s}_2)\in\mathsf{Supp} \atop \bar{s}_2\neq s_2}q({s}_1,\bar{s}_2)\e_{\mc|T(k),X_1(s_1),X_2(s_2)}2^{\im(y;X_1({s}_1),X_2(\bar{s}_2))}&\le&q({s}_1)2^{\im(y;s_1,T(k),X_1(s_1))}.
\end{IEEEeqnarray}
\end{itemize}

In summary, we have the following upper bound on the denominator of \eqref{eq:js102}:
\begin{IEEEeqnarray}{rCl}
\e_{\mc|T(k),X_1(s_1),X_2(s_2)}\sum_{\tilde{s}_1,\bar{s}_2}q(\bar{s}_1,\bar{s}_2)&2&^{\im(y;X_1(\bar{s}_1),X_2(\bar{s}_2))}\le1+q(k)2^{\im(y;k,T(k))}+q(s_2)2^{\im(y;s_2,T(k),X_2(s_2))}\n&&~~~~~~~~~~~~~+q(s_1)2^{\im(y;s_1,T(k),X_1(s_1))}+q({s}_1,{s}_2)2^{\im(y;X_1({s}_1),X_2({s}_2))}.
\end{IEEEeqnarray}
Substituting this in \eqref{eq:js102} gives:
\begin{IEEEeqnarray}{rCl}
\dfrac{q(s_1,s_2)2^{\im(y;X_1(s_1),X_2(s_2))}}{\e_{\mc|T(k),X_1(s_1),X_2(s_2)}\sum_{\bar{s}_1,\bar{s}_2}q(\bar{s}_1,\bar{s}_2)2^{\im(y;X_1(\bar{s}_1),X_2(\bar{s}_2))}}&\ge&\left(q(s_1,s_2)^{-1}2^{-\im(y;X_1({s}_1),X_2({s}_2))}\right.\n&&~+q(s_1,s_2|k)^{-1}2^{\im(y;k,T(k))-\im(y;X_1({s}_1),X_2({s}_2))}\n&& ~+q(s_1|s_2)^{-1}2^{\im(y;s_2,T(k),X_2(s_2))-\im(y;X_1({s}_1),X_2({s}_2))}\n&&\left. +q(s_2|s_1)^{-1}2^{\im(y;s_1,T(k),X_1(s_1))-\im(y;X_1({s}_1),X_2({s}_2))}+1\right)^{-1}\n
&=&\left(2^{h(s_1,s_2)-\im(y;X_1({s}_1),X_2({s}_2))}\right.\n&&~+2^{h(s_1,s_2|k)-\im(y;X_1({s}_1),X_2({s}_2)|k,T(k))}\n&& ~+2^{h(s_1|s_2)-\im(y;X_1({s}_1)|s_2,T(k),X_2(s_2))}\n&&\left. +2^{h(s_2|s_1)-\im(y;X_2({s}_2)|s_1,T(k),X_1(s_1))}+1\right)^{-1}.
\end{IEEEeqnarray}
Using this and the fact that $(T({k}),X_1({s}_1),X_2({s}_2))$  is drawn from $q_Tq_{X_1|S_1T}(.|{s}_1,.)q_{X_2|S_2T}(.|{s}_2,.)=q_{TX_1X_2|S_1,S_2}(.|{s}_1,{s}_2)$, we have:
\begin{IEEEeqnarray}{rCl}
\e\mathsf{P}[C]&\ge&\sum_{s_1,s_2,y}\sum_{t,x_1,x_2}q(s_1,s_2)q(t)q(x_1|s_1,t)q(x_2|s_2,t)q(y|x_1,x_2)\left(2^{h(s_1,s_2)-\im(y;x_1,x_2)}\right.\n&+&\left.2^{h(s_1,s_2|k)-\im(y;x_1,x_2|k,t)} +2^{h(s_1|s_2)-\im(y;x_1|s_2,t,x_2)}+2^{h(s_2|s_1)-\im(y;x_2|s_1,t,x_1)}+1\right)^{-1}\n
&=&\e_{q_{S_1S_2TX_1X_2Y_1Y_2}} \left(1+2^{h(S_1|S_2)-\im(Y;X_1|X_2,S_2,T)}+
                         2^{h(S_2|S_1)-\im(Y;X_2|X_1,S_1,T)}\right.\n&&\left.~~~~~~~~~~~~~~~~~~~~~~~~~~~~~~~+2^{h(S_1,S_2|K)-\im(Y;X_1,X_2|K,T)}+2^{h(S_1,S_2)-\im(Y;X_1,X_2)}\right)^{-1}.
\end{IEEEeqnarray}
This concludes the proof.
\end{IEEEproof}


\begin{thebibliography}{10}
\bibitem{strassen}
V. Strassen, “Asymptotische Absch\"atzungen in Shannon's Informations theorie,” in Trans. Third. Prague Conf. Inf. Theory, 1962, pp. 689–723.
\bibitem{PPV}
Y. Polyanskiy, H. V. Poor, and S. Verd\'u, “Channel coding in the finite blocklength regime,” {\em IEEE Trans. Inf. Theory}, 56(5), 2307 – 2359, 2010.

\bibitem{verdujscc}
V. Kostina, S. Verd\'u, ``Lossy joint source-channel coding in the finite blocklength regime", arXiv:1209.1317, Sep. 2012.

\bibitem{kochman}
D. Wang, A. Ingber, and Y. Kochman, “The dispersion of joint source- channel coding,” in Allerton Conference, 2011, arXiv:1109.6310.
\bibitem{macf}
V. Y. F. Tan and O. Kosut, “On the dispersions of three network information theory problems,” arXiv:1201.3901, Feb 2012.
\bibitem{Tan}
V. Y. F. Tan, ``Transmission of Correlated Sources over a MAC: A Gaussian Approximation-Based Analysis," in Allerton Conference, 2012.
\bibitem{Renner1}
L. Wang and R. Renner, ``One-shot classical-quantum capacity and hypothesis testing ", Physical Review Letters, 2012.
\bibitem{Renner2}
M. Berta, M. Christandl and R. Renner, ``The quantum reverse shannon theorem based on one-shot information theory," Commun. Math. Phys. 306, 579–615, 2011.
\bibitem{verdual}
S. Verd\'u,  ``Non-Asymptotic Achievability Bounds in Multiuser Information Theory", Allerton Conference, Oct. 2012.
\bibitem{watanabe}
S.~Watanabe, S.~Kuzuoka, V.~Y. F. Tan,
``Non-Asymptotic and Second-Order Achievability Bounds for Coding With Side-Information," 	arXiv:1301.6467, Jan 2013.
\bibitem{cuff12}
P. Cuff,
``Distributed channel synthesis," 	arXiv:1208.4415, Aug. 2012.
\bibitem{NIT}
A. El Gamal, Y.-H. Kim, ``Network Information Theory," Cambridge University Press, 2011.
\bibitem{berger}
T. Berger,
``Multiterminal Source Coding,"
Chapter in "The Information Theory Approach to Communications" (G. Longo, editor), Springer-Verlag, 1978.
\bibitem{tung}
S.-Y. Tung,
``Multiterminal source coding", Ph.d. Thesis, Cornell university, NY, 1978.

\bibitem{heegard}
C. Heegard and T. Berger, ``Rate distortion when side information may be absent," {\em IEEE Trans. on Inf. Theory}, 31(6), pp. 727--734,
1985.
\bibitem{kaspi}
A. H. Kaspi, ``Rate-distortion function when side-information may be present at the decoder," {\em IEEE Trans. on Inf. Theory}, 40(6), pp. 2031--2034, 1994.

\bibitem{ZB}
Z. Zhang and T. Berger,
``New results in binary multiple descriptions",  {\em IEEE Trans. on Inf. Theory}, 33(4), pp. 502--521, 1987.
\bibitem{cuffmd}
J. Wang, J. Chen, L. Zhao, P. Cuff and H. permuter,
``On the role of the refinement layer in multiple description coding  and scalable coding",  {\em IEEE Trans. on Inf. Theory}, 57(3), pp. 1443--1456, 2011.
\bibitem{ceg}
T.~Cover, A.~El. Gamal and M.~Salehi,
``￼Multiple Access Channels with Arbitrarily Correlated Sources,"  {\em IEEE Trans. Inf. Theory}, 26(6), 648--657, 1980.

\bibitem{CLT}
V. Bentkus, ``On the dependence of the Berry–Esseen bound on dimension," Journal of Statistical Planning and Inference, 113(2), 385-402, 2003.

\end{thebibliography}
\end{document}